\documentclass[12pt]{article}

\usepackage{times}
\usepackage{times, xcolor, graphicx, graphics, cite, amsmath, upgreek, multicol, gensymb, textgreek}
\usepackage[hidelinks]{hyperref}
\usepackage[pagewise, modulo]{lineno}
\usepackage[labelfont={bf}]{caption}

\newcounter{lastnote}

\begin{document} 
	\sloppy

	\baselineskip24pt

		{\parindent0pt 
		
		\Huge{Microstructural and material property changes in severely deformed Eurofer-97} 
		
		\bigskip
		
		\Large
		{Kay Song $^{\text{a}, \star}$, Guanze He $^{\text{b}}$, Abdallah Reza $^{\text{a}}$, Tamas Ung\'{a}r $^{\text{c}}$, Phani Karamched $^{\text{b}}$, David Yang $^{\text{d}}$, Ivan Tolkachev $^{\text{a}}$, Kenichiro Mizohata $^{\text{e}}$, David E J Armstrong $^{\text{b}}$, Felix Hofmann $^{\text{a}, \dagger}$}
		
			\bigskip
			\large{$^{\text{a}}$ Department of Engineering Science, University of Oxford, Parks Road, Oxford, OX1 3PJ, UK} \\
			\large{$^{\text{b}}$ Department of Materials, University of Oxford, Parks Road, Oxford, OX1 3PH, UK} \\
			\large{$^{\text{c}}$ Department of Materials Physics, E\"{o}tv\"{o}s University,
			Budapest, P.O.Box 32, H‑1518, Hungary} \\
			\large{$^{\text{d}}$ Condensed Matter Physics and Materials Science Department, Brookhaven National Laboratory, P.O. Box 5000, Upton, NY 11973-5000} \\
			\large{$^{\text{e}}$ University of Helsinki, P.O. Box 64, 00560 Helsinki, Finland}
		
		\bigskip
			\large{$^{\star}$Corresponding author email: kay.song@eng.ox.ac.uk} \\
			\large{$^{\dagger}$felix.hofmann@eng.ox.ac.uk}
			
		\bigskip
		\begin{multicols}{2}
			\large{ORCID:}\\
			\normalsize{Kay Song: 0000-0001-8011-3862 \\ Guanze He: 0000-0002-3638-1445 \\ Abdallah Reza: 0000-0001-9594-2591 \\ Tamas Ung\'{a}r: 0000-0003-4392-8856 \\ Phani Karamched: 0000-0002-5599-866X \\ \\ David Yang: 0000-0002-1062-7371 \\ Ivan Tolkachev: 0000-0001-8985-726X \\ Kenichiro Mizohata: 0000-0003-1703-2247 \\ David E J Armstrong: 0000-0002-5067-5108 \\  Felix Hofmann: 0000-0001-6111-339X}
		\end{multicols}
		}
	\newpage
	\begin{abstract}
	Severe plastic deformation changes the microstructure and properties of steels, which may be favourable for their use in structural components of nuclear reactors. In this study, high-pressure torsion (HPT) was used to refine the grain structure of Eurofer-97, a ferritic/martensitic steel. Electron microscopy and X-ray diffraction were used to characterise the microstructural changes. Following HPT, the average grain size reduced by a factor of $\sim$ 30, with a marked increase in high-angle grain boundaries. Dislocation density also increased by more than one order of magnitude. The thermal stability of the deformed material was investigated via in-situ annealing during synchrotron X-ray diffraction. This revealed substantial recovery between 450 K - 800 K. Irradiation with 20 MeV Fe-ions to $\sim$ 0.1 dpa caused a 20\% reduction in dislocation density compared to the as-deformed material. However, HPT deformation prior to irradiation did not have a significant effect in mitigating the irradiation-induced reductions in thermal diffusivity and surface acoustic wave velocity of the material. These results provide a multi-faceted understanding of the changes in ferritic/martensitic steels due to severe plastic deformation, and how these changes can be used to alter material properties.
	
	\end{abstract}
	
	Keywords: High-pressure torsion, Eurofer-97, nanocrystalline microstructure, annealing, ion-irradiation
	\medskip
	
	\newpage
	
	\begin{center}
		\small{\textbf{Graphical Abstract}}
		\begin{figure}[h!]
			\includegraphics[width=0.9\textwidth]{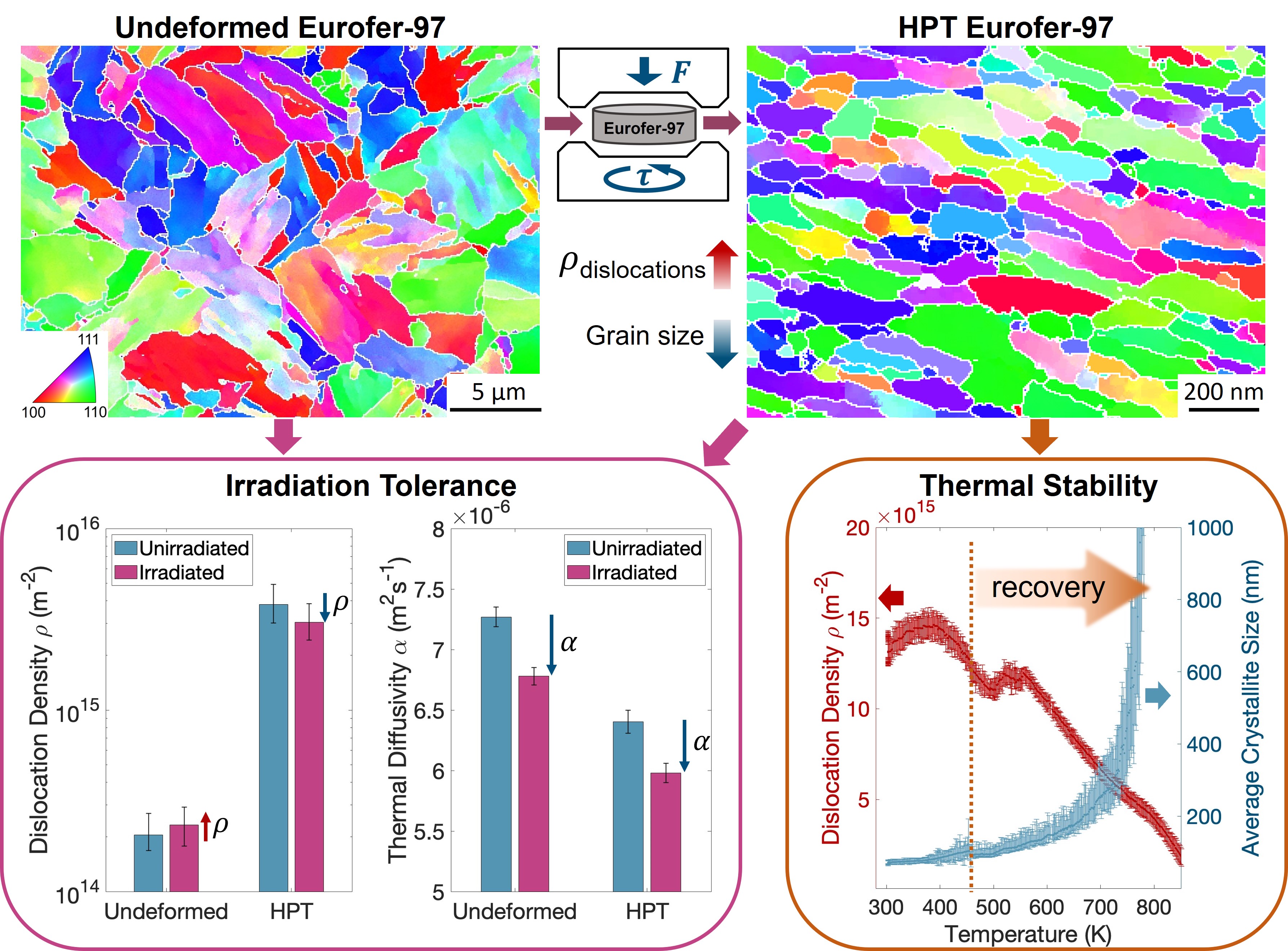}
			\centering
		\end{figure}
	
	\end{center}
	
	\newpage
	
	
	\section{Introduction}
		 A key challenge for next-generation nuclear power is the development and processing of highly radiation-resistant materials for structural components \cite{Moslang2005}. Reduced activation ferritic/martensitic steels, such as Eurofer-97, are leading candidates due to their favourable mechanical and thermal properties \cite{Lindau2005}, as well as good resistance to void swelling \cite{Zinkle2012}. While the exchange of Mo, Nb and Ni for W and Ta in Eurofer-97 helps to achieve low activation capability \cite{Ehrlich1999}, the material still suffers from irradiation-induced changes such as hardening and embrittlement, which affect its performance below 600 K \cite{Klimenkov2011}.
		
		Recently, nanostructuring and grain refinement via severe plastic deformation (SPD) has been proposed as a potential method for increasing the radiation tolerance of materials \cite{Enikeev2019}. The goal for a radiation-resistant material is for its structure and properties to remain stable, or unchanged, under irradiation. It is proposed that SPD increases the density of grain boundaries in the material, which act as sinks for irradiation-induced defects such as vacancies and interstitials, removing them from the bulk material and reducing their overall effect on the material properties \cite{Zhang2018}.
		
		The effects of SPD on the irradiation response of structural steels include reducing irradiation-induced segregation of alloyed solutes \cite{Etienne2011, Radiguet2008}, reducing the population of irradiation-induced defects \cite{Mazilkin2020}, and lowering irradiation hardening \cite{Alsabbagh2014}. Some studies have shown that grain refinement can suppress irradiation-induced void swelling by over an order of magnitude \cite{Gigax2017, Song2014, Sun2015}. However, Aydogan \textit{et al.} have found higher swelling near the surface region (top 1 \textmu m) in the irradiated SPD material compared to the as-annealed material \cite{Aydogan2017}.
		
		The suitability of grain refinement by SPD for increasing radiation resistance of structural steels is also unclear due to the complexity of studying different steels with varying compositions and crystal structures. For example, in face-centred cubic (FCC) materials, the grain refinement mechanism depends greatly on the stacking fault energy and twinning. However for body-centred cubic (BCC) materials, the results of SPD depend more on dislocation slip and the plasticity of the individual material \cite{Cao2018, Edalati2022a}. The range of conditions over which SPD can be carried out (method of deformation, temperature, pre-/post-annealing treatments) also affect the grain refinement and resultant material properties \cite{Enikeev2019, Edalati2022}. Furthermore, the process of SPD itself also changes the properties of steels by increasing yield strength but also reducing ductility, which will affect its operational performance \cite{Valiev2002, Cao2018}. Therefore, more comprehensive studies of the combined effect of SPD and irradiation are warranted in light of its potential use to modify nuclear structural materials. In particular, there are comparatively fewer studies in the literature on the effect of SPD on ferritic/martensitic steels compared to austenitic steels \cite{Cao2018, Edalati2022}. As such, the study of Eurofer-97, a prototypical ferritic/martensitic steel is particularly interesting.
		
		High-pressure torsion (HPT) is a method which subjects materials to a compressive force and simultaneous torsional straining. It is an effective method of producing nanocrystalline materials with grain sizes less than 100 nm \cite{Zhilyaev2008}. HPT produces a disc with radially increasing shear strain, $\gamma$, from 0 at the centre of the disc to:
		\begin{equation}
			\gamma = \frac{2\pi N r}{h}
		\end{equation}
		where $N$ is the number of rotations, $r$ is the distance from the centre of the disc, and $h$ is the thickness of the disc \cite{Zhilyaev2008}. As such, a range of strains can be accessed within the same sample.
		
		Here we present a study on the effect of HPT, performed at room temperature, on the microstructure, thermal stability, and irradiation response of Eurofer-97. A range of experimental characterisation techniques including electron microscopy, X-ray diffraction and transient grating spectroscopy were carried out. The results allow us to study the evolution of grain size, dislocation density and character as a function of deformation, annealing temperature and irradiation. The thermal transport and elastic properties of the deformed and irradiated materials were also probed. Considering the combination of these results, a multi-faceted picture of the effect of HPT on the structure and properties of Eurofer-97, as well as its thermal stability and irradiation response, emerges. This in turn allows us to examine the effectiveness of SPD as a means of enhancing the radiation resistance and material properties of Eurofer-97 as a nuclear structural component. 
		
	\section{Materials and Methods}
		\subsection{Sample Preparation}
		
		The Eurofer-97 material was provided by the UK Atomic Energy Authority. The original 4 mm thick plate was produced by B\"{o}hler Bleche GMBH (Heat 993394) \cite{BohlerBleche2003}. After hot rolling, the material was normalised at 1253 K for 10 minutes and then tempered at 1033 K for 1.5 hours. The manufacturer-provided composition is shown in Table \ref{tab:composition}. 
		
		\begin{table} [h!]
			\begin{center}
				\begin{tabular}{ |c|c|c|c|c|c|c|c|c| }
					\hline
					\textbf{Element} & Cr	&	W	&	V	&	Ta	&	C	&	 Mn	&	N	& Fe \\
					\hline
					\textbf{wt\%} & 9.08	&	1.07	&	0.24	&	0.13	&	0.11	&	0.56	&	0.04 &  Bal.	\\
					\hline
				\end{tabular}
				\caption{The chemical composition of Eurofer-97 with all values in wt\% \cite{BohlerBleche2003}.}
				\label{tab:composition}
			\end{center}
		\end{table}
		
		Samples were cut into discs of 5 mm diameter and 1 mm thickness. HPT was performed using a Zwick Roell Z100 Materials Testing machine in a quasi-constrained set-up \cite{Zhilyaev2008} with custom-made anvils made from M42 tool steel. The samples were compressed with 80 kN of force ($\sim$ 4 GPa pressure), and 9 complete rotations were performed while maintaining this force. The final specimen thickness after HPT was 0.5 mm. Undeformed samples were also retained as reference specimens. Shear strains investigated for the specimens in this study range from 0 at the centre of the disc to 280 at the edge of the disc.
		
		All samples were mechanically ground with SiC paper, then polished with diamond suspension and colloidal silica (0.04 \textmu m). The final polishing step was electropolishing with 5\% perchloric acid and 15\% ethylene glycol monobutyl ether in ethanol at 293 K, using a voltage of 30 V for 2-3 minutes using a Struers LectroPol-5.
		
		\subsection{Ion Irradiation}
\begin{figure}[h!]
			\centering
			\includegraphics[width=0.6\textwidth]{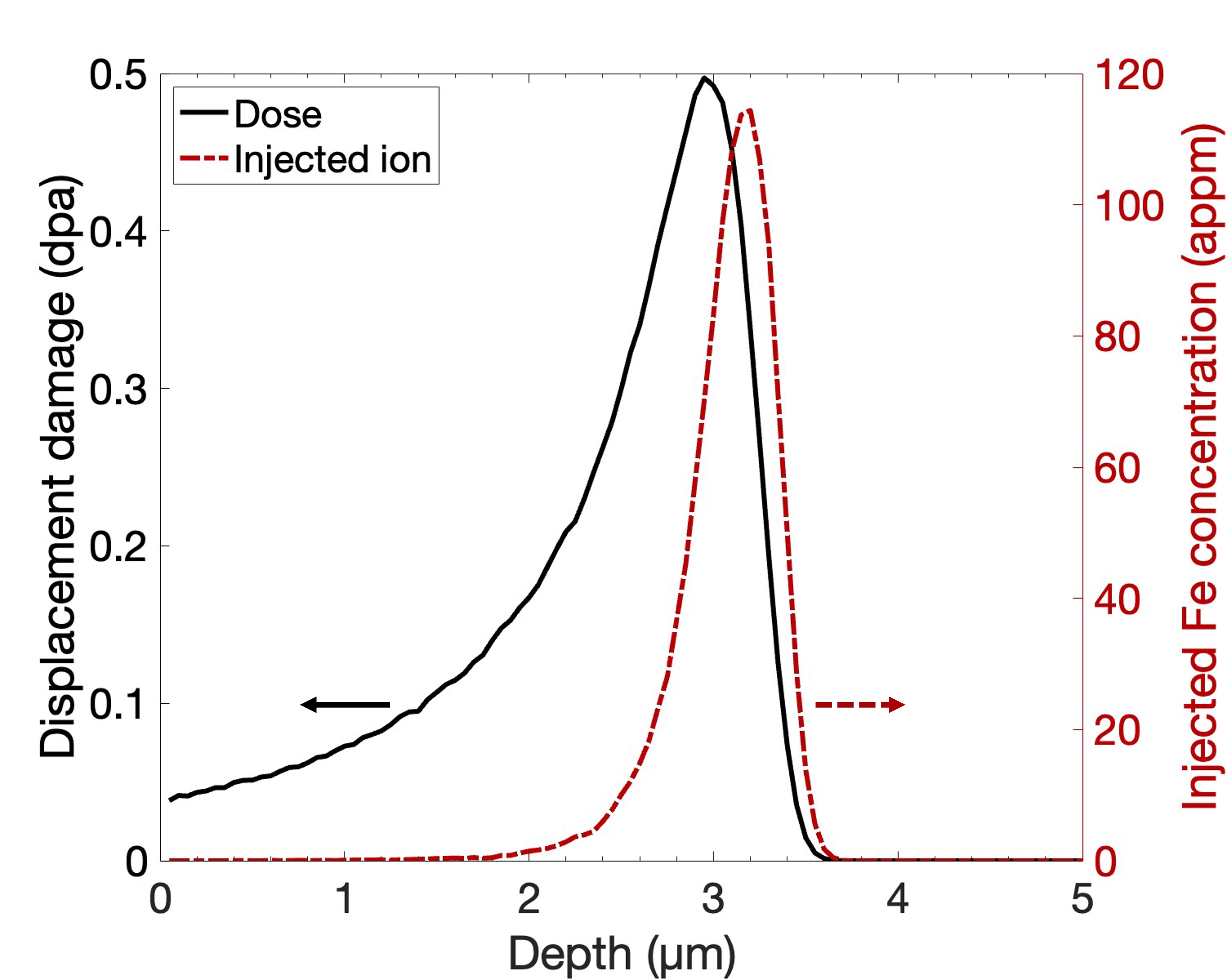}
			\caption{The calculated dose and injected ion profile from SRIM \cite{Ziegler2010} for 20 MeV Fe-ion irradiation on Fe target. }
			\label{fig:dpa}
		\end{figure}

		Irradiation with 20 MeV Fe$^{4+}$ ions at room temperature was performed using the tandem accelerator at the Helsinki Accelerator Laboratory. Both the HPT-processed and undeformed samples were irradiated. The dose profile was calculated using the Quick K-P model in the SRIM code \cite{Ziegler2010}, using 20 MeV Fe ions on a Fe target with 40 eV displacement energy at normal incidence (Figure \ref{fig:dpa}). The calculated damaged layer extends to 3.5 \textmu m below the sample surface. The average dose in the first 2 \textmu m below the surface, where the injected ion concentration is small, is 0.08 displacements-per-atom (dpa). The peak dose is 0.5 dpa at a depth of 3 \textmu m and the average dose across the whole implanted layer (from 0 to 3.5 \textmu m depth) is 0.17 dpa. All these doses fall into the transition regime before the onset of defect microstructure saturation in ferritic materials \cite{Derlet2020, Song2020, Song2023}. 
		
		\subsection{Electron Microscopy}
		SEM-based characterisation was carried out using a Zeiss Merlin FEG-SEM at the David Cockayne Centre for Electron Microscopy, University of Oxford. Backscattered electron (BSE) imaging was used for qualitative characterisation of grain morphology and carbide distribution on the sample surface. For quantitative grain size analysis of the undeformed material, electron backscatter diffraction (EBSD) was carried out at 30 kV with a beam current of 10 nA.
		
		For the HPT-processed material, conventional EBSD did not provide sufficient spatial resolution for microstructural characterisation due to the refined grain size. Lift-out samples were made using focused ion beam (FIB) milling (Zeiss Crossbeam 540 FIB-SEM and Zeiss NVision 40 FIB-SEM). The samples were thinned with Ga ions to approximately 100 nm. Lift-outs were taken at 2 different locations on each HPT sample (unirradiated and irradiated), corresponding to shear strains of $\gamma = 110$ and $\gamma = 230$.
		
		TEM imaging was carried out with a Jeol 2100 TEM with a LaB$_{\text{6}}$ source, operating at 200 kV. EDX measurements of the TEM samples to characterise carbides in the material were carried out on a JEOL ARM-200F TEM, operating at 200 kV. On-axis transmission Kikuchi diffraction (TKD) measurements were carried out using a Zeiss Merlin FEG-SEM at 30 kV, for quantitative characterisation of grain size and dislocation density.
		
		Grain size analysis was carried out using the Oxford Instrument HKL Channel 5 Tango software, with the grain boundary critical misorientation set to 10$^{\circ}$. Geometrically necessary dislocation (GND) analysis was carried out in the same software, which uses the kernel average misorientation approach described in \cite{Calcagnotto2010}. The data was first filtered with the Kuwahara method to remove some noise due to the small step sizes used for TKD \cite{Wright2015}. The average misorientation at each point was calculated with respect to its second nearest neighbours as it was determined to be optimum for resolving the GND density distribution while not being too affected by step size-related noise.
		
		\subsection{Synchrotron X-ray Diffraction}
		\subsubsection{Grazing-incidence X-ray Diffraction}
		X-ray diffraction (XRD) measurements were made at the Diamond Light Source I11 beamline with high-resolution multi-analyser crystal (MAC) detectors. The detector set-up consists of 5 MAC arms, which hold 9 analysing Si crystals each, mounted on a 2$\uptheta$-circle with 30$^{\circ}$ separation \cite{Thompson2009, Tartoni2008}. Due to the narrow rocking curves of the Si crystal, a high instrument resolution can be achieved. A grazing-incidence geometry was used with a 10$^{\circ}$ incidence angle, and X-ray energy of 15 keV ($\lambda$ = 0.82656 \r{A}, $\Delta$E/E = $10^{-4}$). This ensured that the majority of the diffraction signal was dominated by contributions from the top 3 - 4 \textmu m thick layer of the samples, where the implanted layer lies. The beam size was approximately 250 \textmu m in height and 1 mm in width. Measurements were taken at positions halfway between the centre and edge of the disc samples. Accounting for the size of the beam and its footstep on the sample surface due to the incident angle, the area measured corresponds to shear strains between 65 to 160.
		
		\subsubsection{In-situ heating X-ray Diffraction}
		To study the thermal stability of the deformed material, in-situ heating during XRD was performed at the Diamond Light Source I12 beamline. Samples were mounted in a Linkam TS1500 heating stage for diffraction in transmission geometry. The nominal heating rate was 20 K/min, then holding for 5 minutes at the target temperature, followed by cooling to room temperature at a nominal rate of 30 K/min. The samples were held under an argon atmosphere throughout the heating cycle to prevent oxidation. The highest nominal temperature reached during the annealing cycle was 1323 K. 
				
		The X-ray beam was monochromated to an energy of 80 keV ($\lambda$ = 0.15498 \r{A}, $\Delta$E/E = $10^{-4}$). The energy calibration and the instrument broadening function were measured using a ceria standard sample. The beam size of 0.5 mm $\times$ 0.5 mm at the sample was defined using slits. The position at which measurements were taken on the sample corresponded to a shear strain of $\sim 110$. Two-dimensional XRD patterns were collected on a Pilatus 2M CdTe detector positioned at a distance of 750 mm from the sample. Patterns were collected with 5 seconds of exposure for the entire duration of the heating and cooling cycle. Azimuthal integration of the 2D raw data to obtain intensity vs. 2$\uptheta$ (scattering angle) spectra was performed with the DAWN software package \cite{Filik2017}.
		
		Though the Linkam heating stage provided a temperature reading for each diffraction pattern, a calibration was necessary to determine the actual temperature of the sample due to imperfections in the thermal contact between the stage and the samples. The sample temperature was estimated by fitting the lattice parameter of each diffraction pattern and using the thermal expansion coefficient of Eurofer-97 \cite{Paul2006}. Further details are included in Appendix A. All temperatures discussed from this point onwards refer to the calibrated sample temperature.

		\subsubsection{Convolutional Multiple Whole Pattern (CMWP) Analysis }
		Analysis of the intensity vs. 2$\uptheta$ diffraction spectra (from both I11 and I12 beamlines) was performed using the convolutional multiple whole profile (CMWP) fitting method \cite{Ribarik2019, Ribarik2020}. This allowed for the extraction of crystallite size distribution, dislocation density and dislocation character. Briefly, CMWP analysis is based on fitting the sum of a background spline function and a profile function, which is a convolution of the theoretical size and strain profiles, as well as the measured instrument profile. The crystallite size profile is modelled as a log-normal distribution. The strain profile considers the dislocation density, as well as the characteristic lengthscale of their strain fields, which depends on their physical ordering and arrangement \cite{Wilkens1970, Ungar1999a}.
		
		\subsection{Transient Grating Spectroscopy}
		Transient grating spectroscopy (TGS) is a laser-based non-destructive technique to measure thermal diffusivity and surface acoustic wave velocity. The details of the technique are described elsewhere \cite{Reza2020b, Kading1995, Dennett2017}. Briefly, two short pump laser pulses (0.5 ns duration at 532 nm, with pulse energy $\sim$ 1.5 \textmu J and 1 kHz repetition rate) are overlapped on the sample surface at a fixed angle to form an interference pattern with wavelength $\lambda$. Partial absorption of the light and subsequent thermal expansion occurs at positions of constructive interference, creating a spatial grating on the surface. At the same time, this rapid thermal expansion launches two monochromatic, counter-propagating surface acoustic waves (SAW). Following excitation by the pump beams, this `transient grating' will decay due to the diffusion of thermal energy from maxima to minima and also into the material bulk. By measuring this decay via the diffraction of a set of probe laser beams (continuous wave at 559.5 nm), the thermal diffusivity and the SAW velocity of the material can be determined. The key strength of TGS applied to ion-irradiated materials is that the thickness of the probed layer is approximately equal to $\tfrac{\lambda}{\pi}$, so $\lambda$ can be adjusted to ensure the signal is dominated by contributions from the irradiated layer.
		
		This study uses $\lambda = 5.116 \pm 0.002$ \textmu m for most measurements. Extra measurements were made for the HPT-deformed and irradiated sample with $\lambda = 5.116 \pm 0.002$ \textmu m. The resultant probed layer thickness for all measurements was $\sim$ 1.6 \textmu m, which ensures that the signal originates from the 3.5 \textmu m thick damaged layer in the irradiated samples. The pump and probe beam sizes were respectively 140 \textmu m and 90 \textmu m ($\tfrac{1}{e^{2}}$ width). In order to measure properties corresponding to different shear strains, lines of measurement points spaced 150 \textmu m apart were taken across the diameter of the HPT-processed samples.
		
	\section{Results}
		
		\subsection{Electron Microscopy} \label{sec:em}
		
		\begin{figure}[h!]
			\centering
			\includegraphics[width=0.9\textwidth]{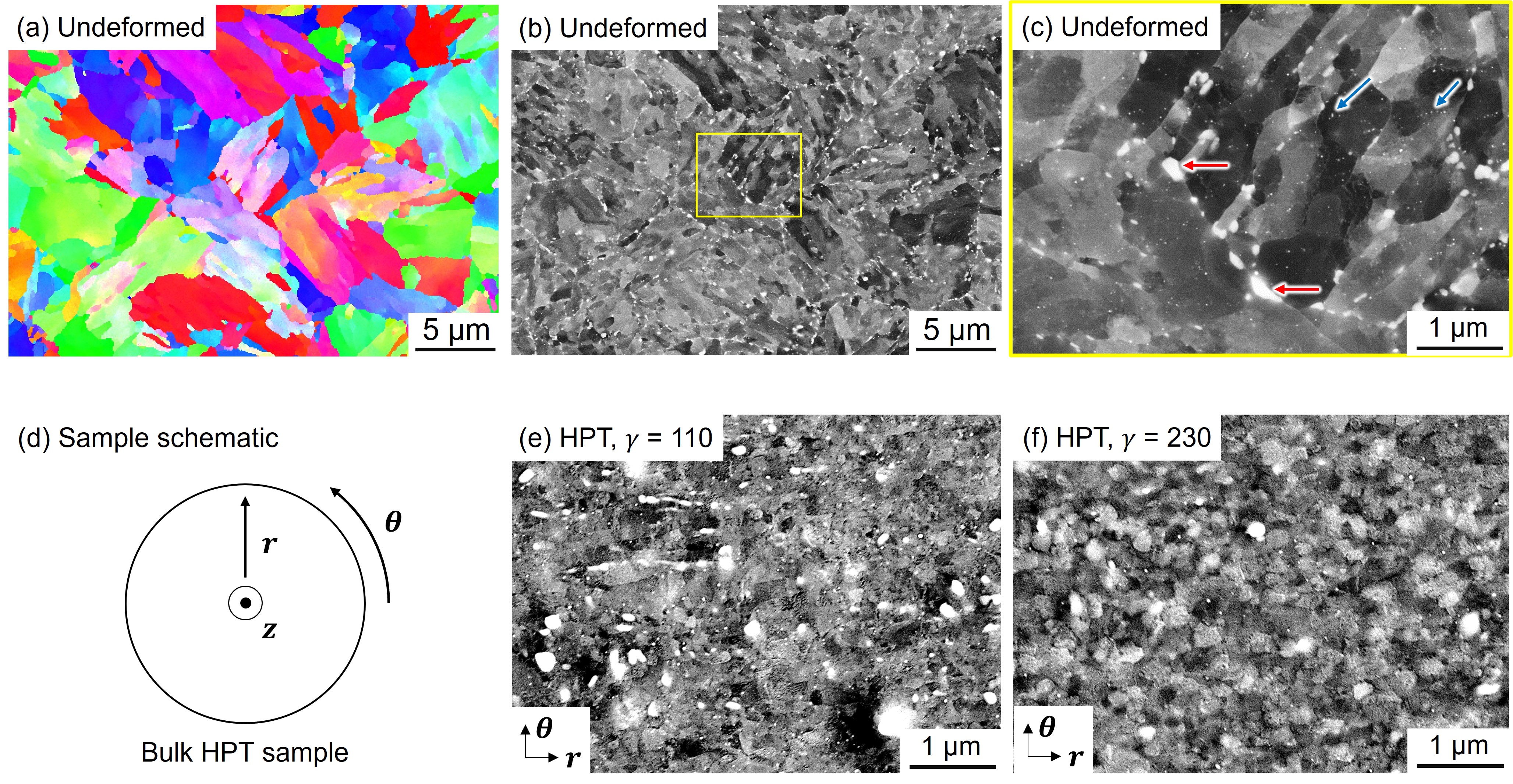}
			\caption{The microstructure of the same area on the as-received Eurofer-97 characterised by (a) EBSD and (b) backscattered electron (BSE) imaging. (c) A higher magnification BSE image of the as-received material (same area enclosed by the yellow rectangle in (b)). Red and blue arrows indicate the presence of large and small carbides, respectively. (d) A schematic of the sample coordinates with respect to the deformed material. The orientation shown applies to all micrographs in this figure. BSE imaging of Eurofer-97 after deformation at (e) $\gamma$ = 110 and (f) $\gamma$ = 230. }
			\label{fig:SEM}
		\end{figure}
		
		EBSD of the as-received Eurofer-97, before deformation via HPT, shows lath structures characteristic of the martensitic microstructure of the material (Figure \ref{fig:SEM}(a)). The area-weighted average grain size of the undeformed material is 5.3 $\pm$ 3.2 \textmu m, using an angular misorientation threshold of 10$^{\circ}$. Following electropolishing, a dispersion of particles, each between a few nm to 200 nm in size, becomes evident on the sample surface (Figure \ref{fig:SEM}(b) and (c)). Particles of similar size and spatial distribution have been observed in other studies of Eurofer-97 and similar steels, and have been identified as carbides \cite{Ganeev2018, Arredondo2020, Aydogan2017}. The larger carbides mainly appear at the boundaries of prior austenite grains and are up to 200 nm in size (red arrows in Figure \ref{fig:SEM}(c)). There are also smaller carbides, $<$ 10 nm in size, distributed more homogeneously across the sample (blue arrows in Figure \ref{fig:SEM}(c)).
		
		In this work, cylindrical coordinates will be used to describe directions in the HPT-deformed material (Figure \ref{fig:SEM}(d)). $\boldsymbol{z}$ is normal to the surface of the HPT sample (outwards positive), $\boldsymbol{r}$ is in the radial direction, and $\boldsymbol{\theta}$ is the tangential direction along which torsion is applied. Following deformation via HPT, a significant level of grain refinement is achieved (Figure \ref{fig:SEM}(e) and (f)). Quantitative measurements of the grain size following HPT will be presented in the subsequent sections as EBSD was not able to achieve the required spatial resolution on the sample surface. From SE and BSE imaging, it can be seen that most of the refined grains are equiaxed, with no preferential alignment to the direction of torsion when observed along the $\boldsymbol{-z}$ direction (in the $\boldsymbol{r}$-$\boldsymbol{\theta}$ plane, Figure \ref{fig:SEM}(e) and (f)). The carbides are still present following HPT but are more homogeneously distributed. A mix of large and small carbides is present with similar size distributions to the undeformed sample.
		
		\begin{figure}[h!]
			\centering
			\includegraphics[width=0.95\textwidth]{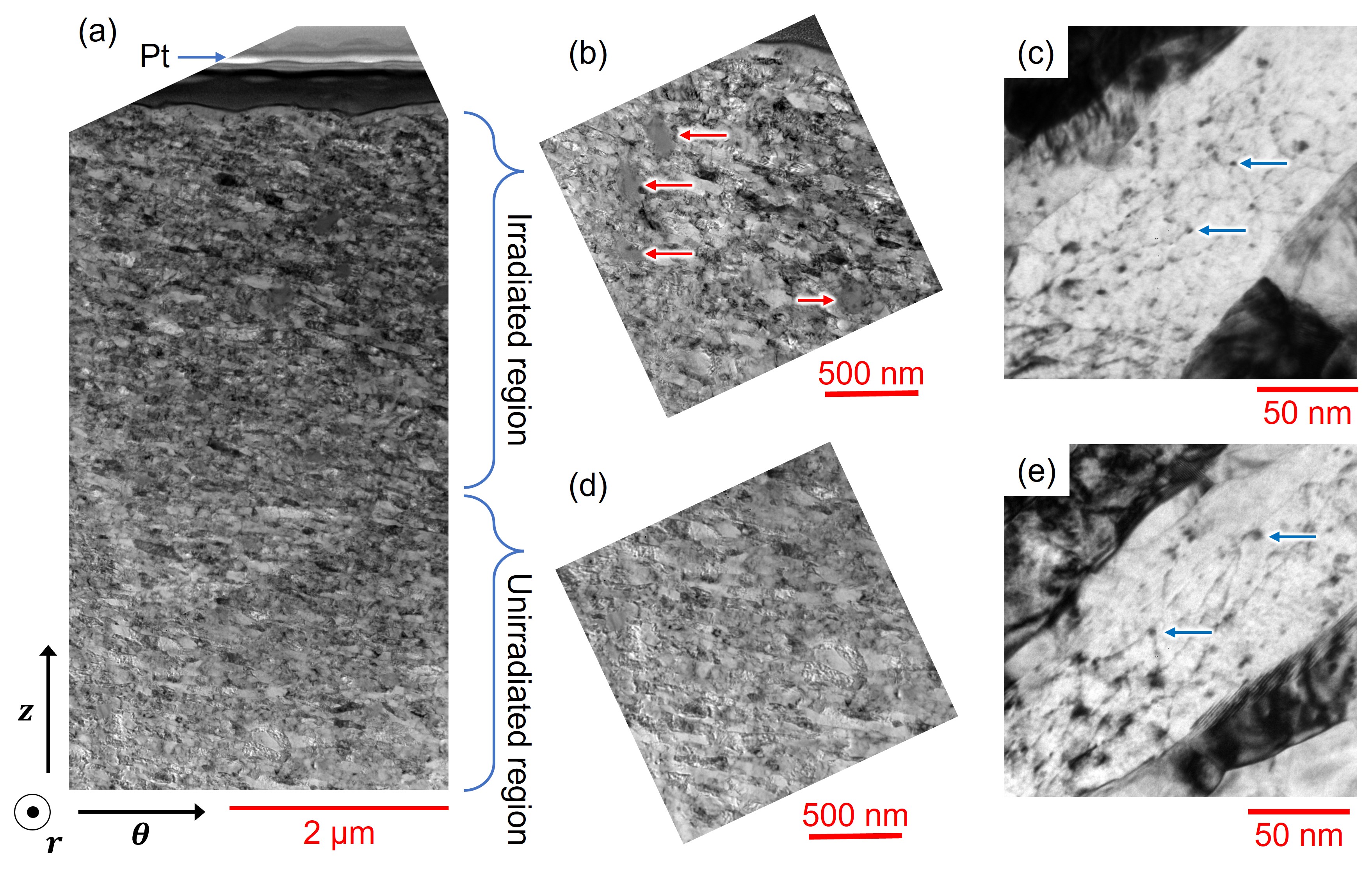}
			\caption{TEM images of the deformed irradiated sample with the lift-out taken from a location corresponding to $\gamma = 230$. (a) Low magnification overview of the lift-out, showing the platinum (Pt) protective layer on the bulk sample surface, with a carbon layer underneath (black). (b)\&(d) Magnified micrographs showing the grain shapes and microstructure in the irradiated and unirradiated regions, respectively. The figures have been rotated to align with the $\boldsymbol{\theta}$-$\boldsymbol{z}$ plane of the bulk sample. (c)\&(e) Bright-field images of individual grains within the irradiated and unirradiated regions respectively. }
			\label{fig:TEM}
		\end{figure}
		
		To further study the grain microstructures of the HPT material, FIB was used to prepare TEM lamella from different samples and radial positions. A full comparison of TEM micrographs from all samples is provided in Appendix B. There is no discernible difference in microstructure between $\gamma = 110$ and $\gamma = 230$, suggesting that strain saturation is reached before $\gamma = 110$. TEM of both the unirradiated and irradiated HPT material also shows that there is no depth dependence of the microstructure down to 10 \textmu m below the bulk sample surface. An example is shown in Figure \ref{fig:TEM}(a) for the top 6 \textmu m of the deformed and irradiated sample. 
		
		When viewed along the radial direction (on the $\boldsymbol{\theta}$-$\boldsymbol{z}$ plane), the grains appear elongated (Figure \ref{fig:TEM}(a)-(c)). The axis of elongation is not completely aligned with the direction of torsion $\boldsymbol{\theta}$ (parallel to the platinum layer on the top edge of the lift-out) but inclined approximately 15$^{\circ}$. This has been observed previously in other studies of HPT materials, even at shear strains beyond microstructural saturation \cite{Hafok2008, Pippan2010, Rathmayr2013, Naghdy2017}. 
		
		The presence of large carbides is detected in the TEM images as indicated by the red arrows in Figure \ref{fig:TEM}(b), similar to those indicated with red arrows in Figure \ref{fig:SEM}(c). EDX was performed to confirm that these regions, which appear grey and without sharp edges, are indeed carbides enriched with chromium and tungsten (Appendix C).
		
		Bright-field images of individual grains were recorded from both the irradiated (Figure \ref{fig:TEM}(c)) and unirradiated (Figure \ref{fig:TEM}(e)) parts of the lift-out. Numerous dislocation lines are evident in both grains. The black spots (blue arrows), most likely carbides, also appear along the length of the dislocation lines. No evidence of irradiation-induced microstructural changes (e.g. dislocation loops) were observed.
		

		\begin{figure}[h!]
			\centering
			\includegraphics[width=0.95\textwidth]{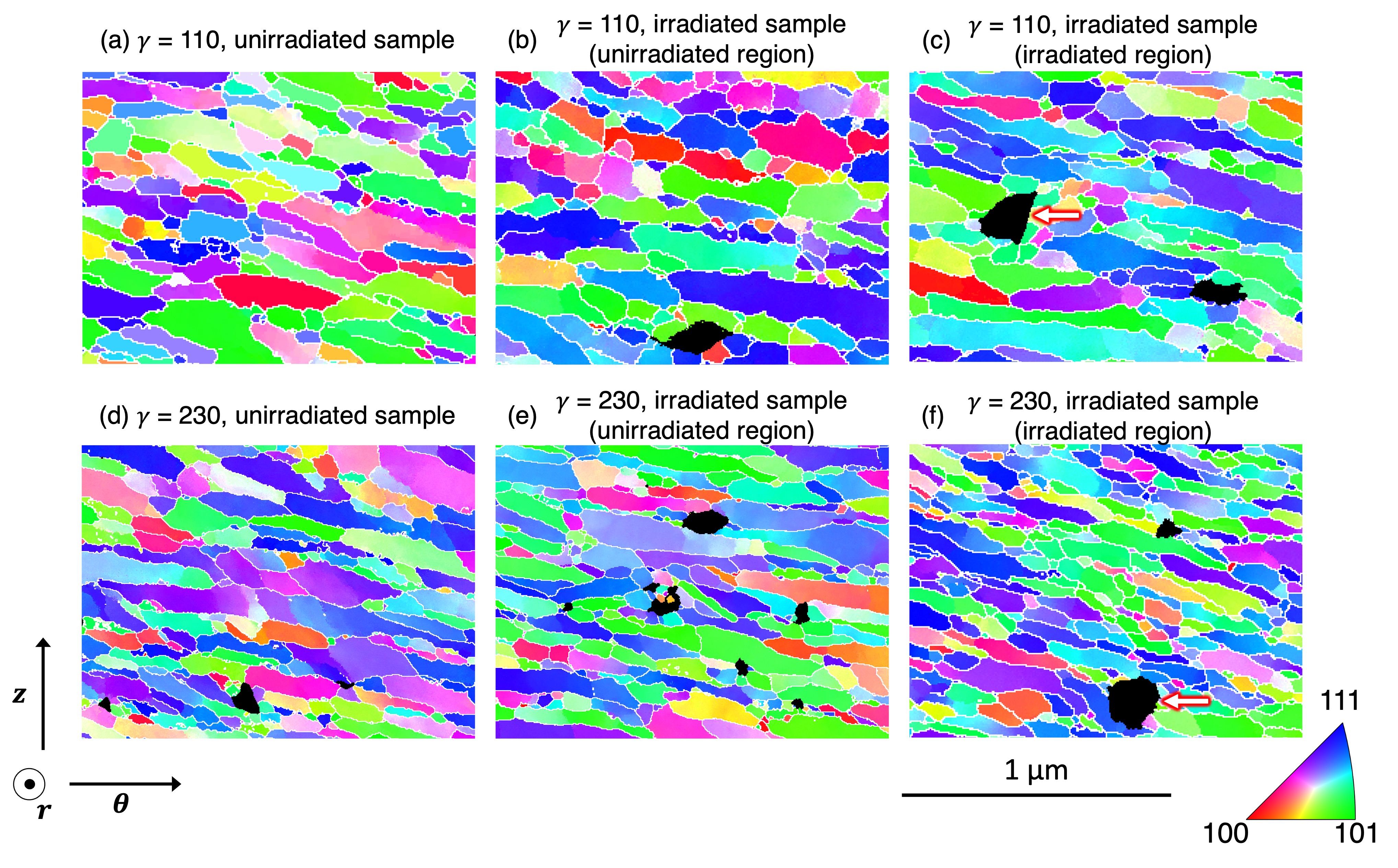}
			\caption{TKD maps taken from the (a)\&(d) unirradiated samples and the irradiated samples in the (b)\&(e) unirradiated regions, and (c)\&(f) irradiated regions, from respectively $\gamma = 110$ and $\gamma = 230$. Note that the top and bottom edges of each sample are parallel to the surface of the bulk sample from which the lift-outs were taken. The grain boundaries (angular misorientation $>$ 10$^{\circ}$) are indicated in white. Non-indexed areas are indicated in black.}
			\label{fig:TKD}
		\end{figure}
		
		Comparisons of the microstructure of different deformed samples were done with TKD measurements (Figure \ref{fig:TKD}). The average indexing rate was $>$ 80\%. Regions of low confidence indexing occurs near the grain boundaries and around carbides (large compact black areas indicated by white arrows with red outlines in Figure \ref{fig:TKD}(c) and (f)). 
		
		Comparing the effect of shear strain on the microstructure (top row compared with bottom row in Figure \ref{fig:TKD}), it can be seen that the aspect ratio of the grains appears slightly larger for $\gamma = 230$ compared to $\gamma = 110$. 
		The difference in strain level does not affect the indexing rate.
		
		Comparing the unirradiated sample with the unirradiated regions of the irradiated sample at each strain level (Figure \ref{fig:TKD}(a) and (c), (b) and (d)), it can be seen that the microstructure at nominally the same deformation conditions is similar across different samples. This confirms the reproducibility of the HPT method used for this study.
		
		There appears to be no discernible irradiation-induced changes to the microstructure in the TKD maps (Figure \ref{fig:TKD}(c) and (f)). The overall grain shapes and distribution of unindexed regions appear to be the same for the irradiated and unirradiated regions of the sample at each strain level. 
		
		\subsubsection{Grain Size}\label{sec:gs}
		\begin{figure}[h!]
			\centering
			\includegraphics[width=0.95\textwidth]{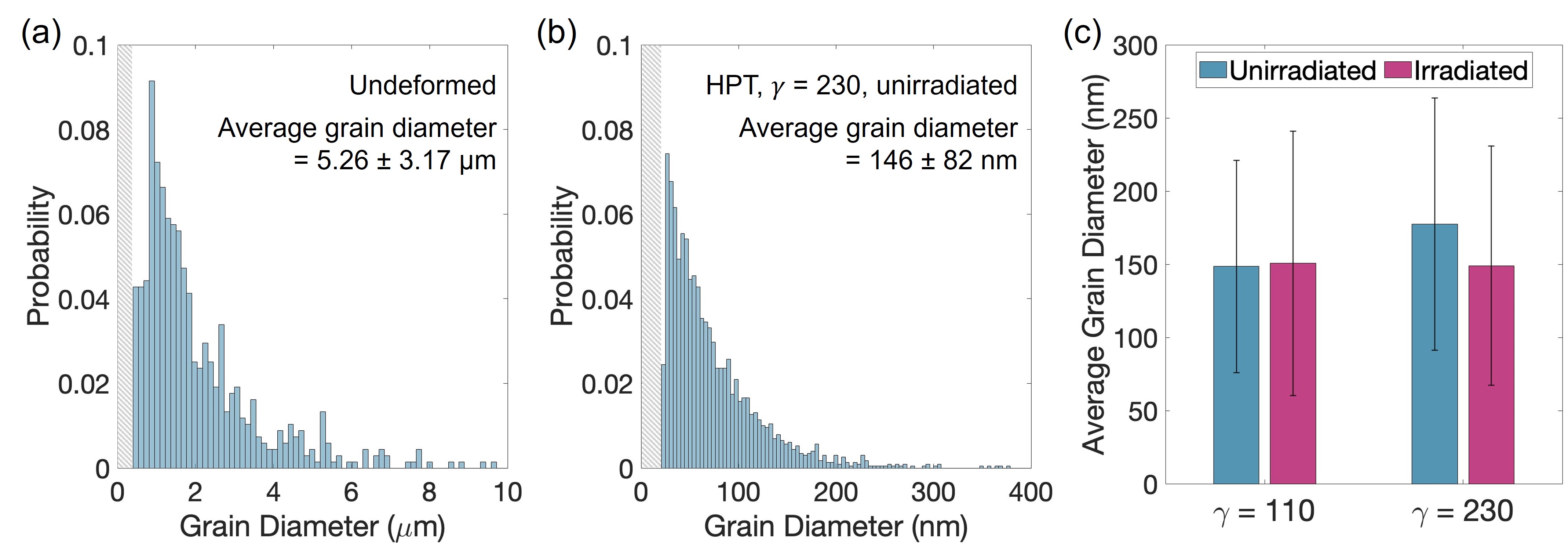}
			\caption{The grain diameter distribution for the (a) undeformed and (b) deformed ($\gamma$ = 230) material from TKD measurements. The grey regions indicate the range of inaccessible grain sizes due to the limited measurement step size. (c) The average grain diameter for the unimplanted and implanted regions at the two strain levels where lift-outs were taken. The error bars represent $\pm$ 1 standard deviation of the grain diameter. Note that all average and standard deviation grain sizes reported are weighted by grain area. }
			\label{fig:grainsize}
		\end{figure}
		
		The grain size distribution shape appears to be similar before and after deformation with HPT (Figure \ref{fig:grainsize}(a) and (b)). Due to the presence of grains much smaller than the step sizes of EBSD and TKD, it was difficult to probe the lower end of the distribution ($<$ 250 nm for EBSD and $<$ 20 nm for TKD). 
		The average grain size following HPT, to a shear strain of $\gamma$ = 230, is reduced by over a factor of 35 compared to the undeformed reference. Comparing the grain sizes following HPT, there does not appear to be much change with strain level within measurement error (Figure \ref{fig:grainsize}(c)). This confirms that the material microstructure is in the saturated regime at or before $\gamma$ = 110. There is also no significant change in grain size following irradiation.
		
		
		\subsubsection{GND Density}
		\begin{figure}[h!]
			\centering
			\includegraphics[width=0.95\textwidth]{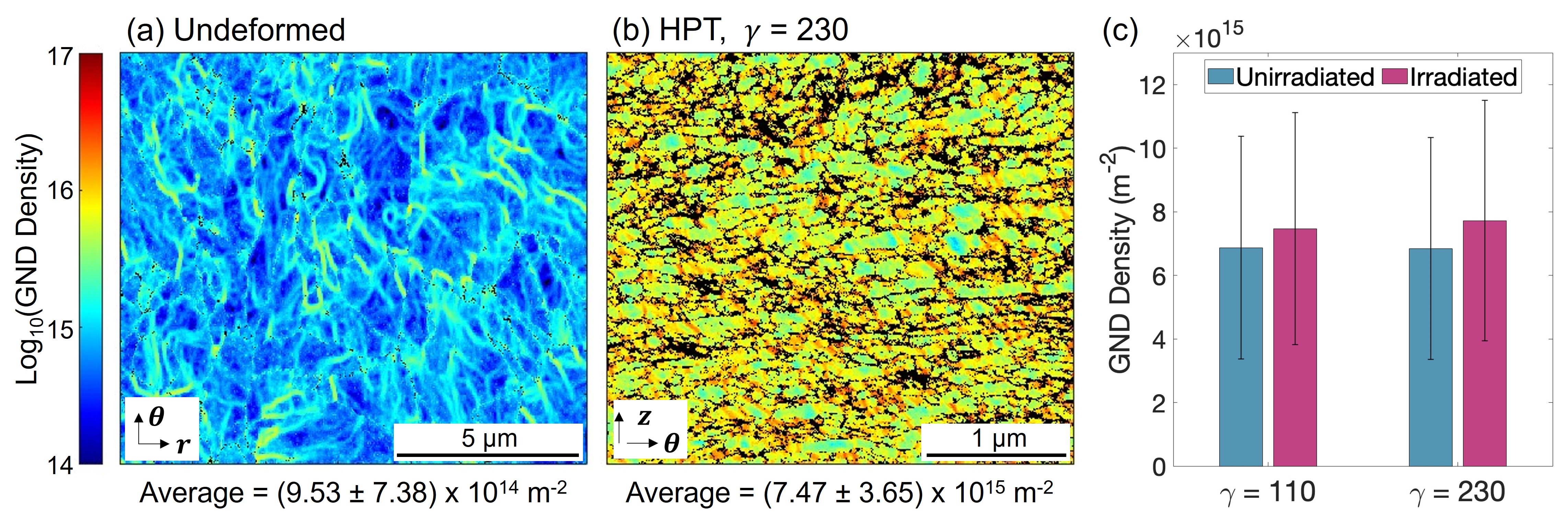}
			\caption{GND density map of the (a) undeformed and (b) deformed material ($\gamma$ = 230). Note that the magnification and sample orientations are not the same for the two samples. (c) The average GND density for the unimplanted and implanted regions at $\gamma$ = 110 and $\gamma$ = 230. The error bars represent $\pm$ 1 standard deviation of the GND density. }
			\label{fig:gnd}
		\end{figure}
		
		The distribution of geometrically necessary dislocations (GND) in the undeformed material shows extended swirling structures that are distributed evenly across and within the grains (Figure \ref{fig:gnd}(a)). The average GND density is on the order of $10^{14}$ - $10^{15}$ m$^{-2}$ as expected for a material that has undergone cold working \cite{Raabe2014, Mazilkin2020}. The GND density increases by an order of magnitude following HPT to $10^{15}$ - $10^{16}$ m$^{-2}$ (Figure \ref{fig:gnd}(b)). The regions of high dislocation density are mainly near grain boundaries. Some bands of high dislocation density are oriented perpendicular to the elongated axis of the grains.
		
		Comparing dislocation density between the different lift-out samples of the deformed materials reveals little difference in average GND density as a function of strain between $\gamma$ = 110 and $\gamma$ = 230 (Figure \ref{fig:gnd}(c)). Once again this confirms that strain saturation is achieved before $\gamma$ = 110. Implantation also did not change the average GND density of the samples. This is unsurprising as at this damage level, most irradiation defects likely exist as point defects or small loops \cite{Yao2008} and thus do not contribute to the large-scale lattice curvature from which GND density is determined.

		\subsubsection{Grain Boundaries and Texture}
		\begin{figure}[h!]
			\centering
			\includegraphics[width=0.95\textwidth]{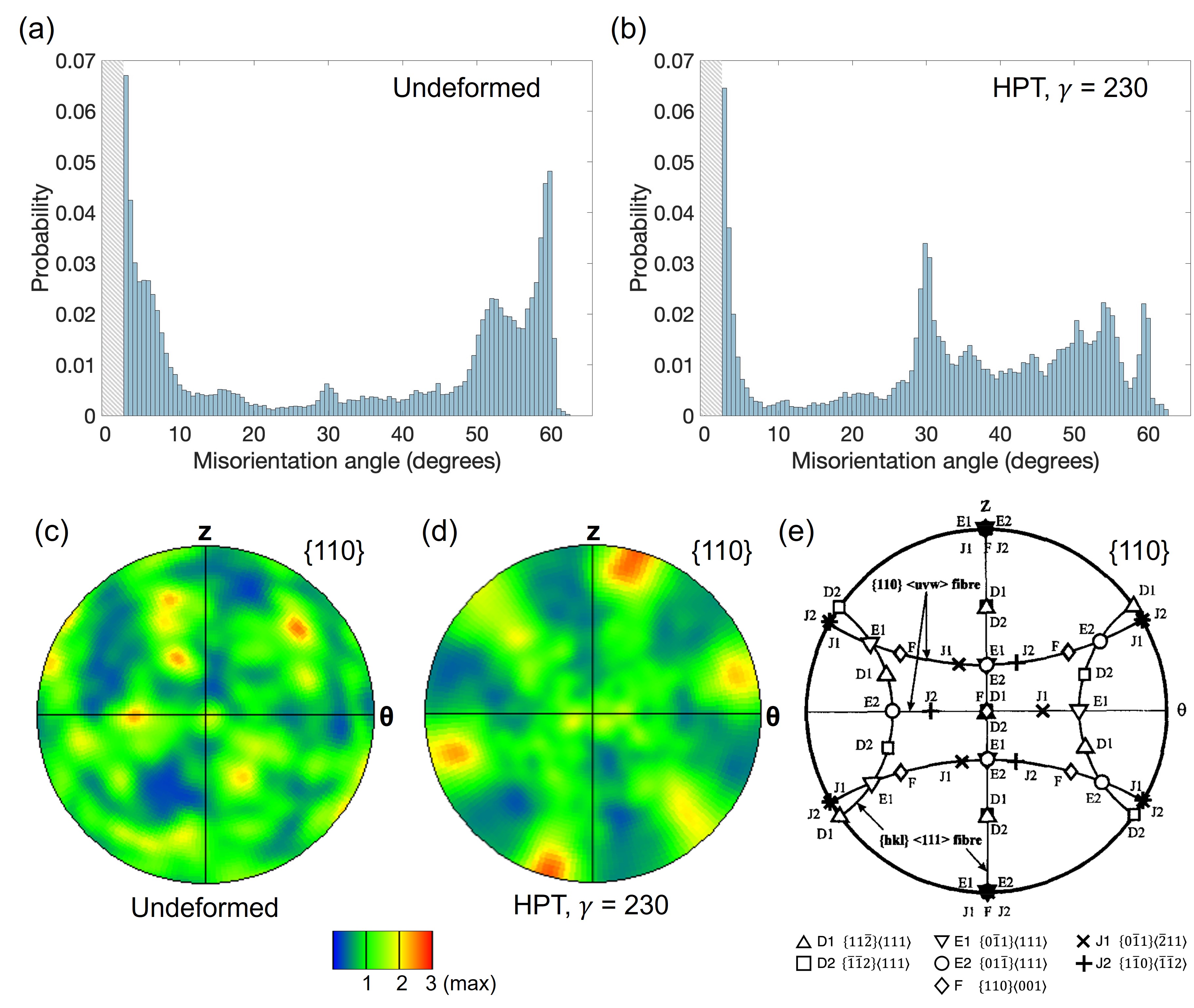}
			\caption{The distribution of grain boundary misorientation angle for the (a) undeformed and (b) deformed material, along with (c)-(d) the corresponding $\{110\}$ pole figures. (e) The theoretical $\{110\}$ pole figure of pure iron that has been processed by torsion \cite{Baczynski1996}.}
			\label{fig:texture}
		\end{figure}
		
		The undeformed Eurofer-97 material has a high proportion of grain boundaries with $<10^{\circ}$ and $>50^{\circ}$ misorientation (Figure \ref{fig:texture}(a)). After torsion, this distribution changes to a large proportion of grain boundaries between $30^{\circ} - 60^{\circ}$ (Figure \ref{fig:texture}(b)). There is a significant peak around $30^{\circ}$ which was greatly enhanced following deformation with HPT. The true proportion of angular misorientation below 2.5$^{\circ}$ could not be accurately determined due to experimental limitations (grey regions in Figure \ref{fig:texture}(a) and (b)).
		
		The $\{110\}$ pole figure for the undeformed Eurofer-97 shows little preferred orientation or texture (Figure \ref{fig:texture}(c)). After deformation with HPT, significant texture evolution occurs and preferred orientation is evident (Figure \ref{fig:texture}(d)). The experimental observations qualitatively agree with the theoretically predicted preferred orientations for torsioned iron (Figure \ref{fig:texture}(e)). There is a $\sim$ 15$^{\circ}$ rotation between the experimentally determined and the predicted texture. This is similar to the angular difference between the top surface of the bulk HPT sample and the axis of elongation for the grains in the $\boldsymbol{\theta}$-$\boldsymbol{z}$ plane (Figure \ref{fig:TKD}(d)).

		\subsection{X-Ray Diffraction} 
		
		\subsubsection{Grazing-Incidence X-Ray Diffraction} \label{sec:i11}
		
		\begin{figure}[h!]
			\centering
			\includegraphics[width=\textwidth]{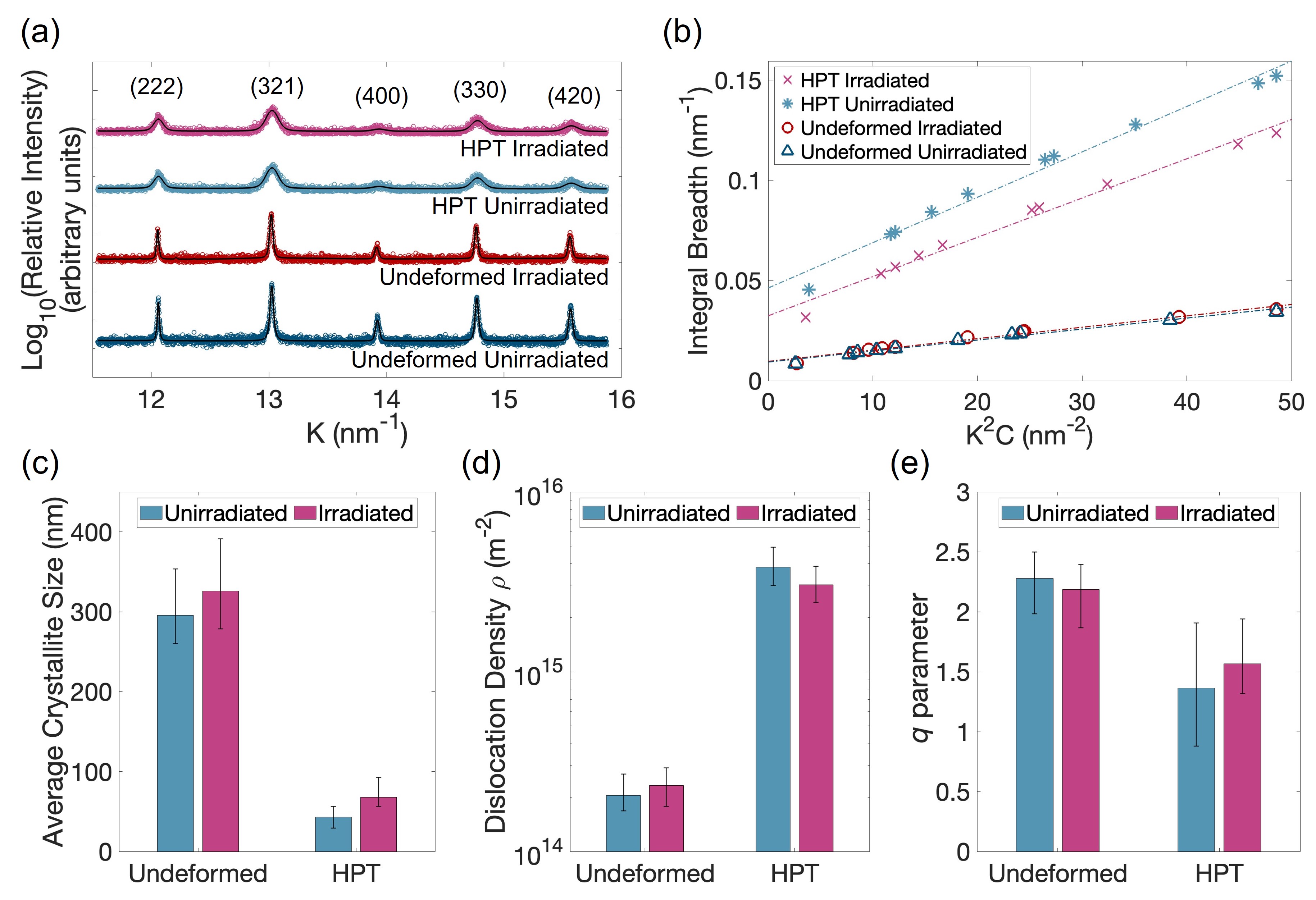}
			\caption{(a) A section of the diffraction patterns, plotted as a function of $K = 2\sin\theta/\lambda$, for the samples investigated. The traces have been vertically offset for clarity. The scatter points are the experimental data points and the black lines are the fitted patterns from CMWP. (b) The modified Williamson-Hall plot with $C$ calculated from CMWP fitting. The dashed lines are included as a guide to the eye. (c) The area-weighted average crystallite size, (d) dislocation density, and (e) $q$ parameter determined from CMWP. Further information about each parameter is given in the main text. The error bars represent the fitting errors calculated by considering the variations in the weighted-sum-of-squares residuals \cite{Ribarik2019}.}
			\label{fig:i11}
		\end{figure}
		
		Peak broadening is clearly observed in the raw diffraction patterns following HPT to $\gamma = 110$ (Figure \ref{fig:i11}(a)). This suggests an increase of intragranular strain and/or a decrease in grain size. The diffraction patterns shown in Figure \ref{fig:i11}(a) are functions of $K = 2\sin\theta/\lambda$, and the fittings from CMWP (overlaid black lines) appear to agree well with the measured data. Due to strain anisotropy, the peak broadening is not isotropic, i.e. it is not linear with $K$. Strain anisotropy in peak broadening can be rectified by taking into account the dislocation contrast factors, $C$, in the modified Williamson-Hall plot \cite{Ungar1996}, where the FWHM or the integral breadth is plotted against $K^{2}C$. The contrast factor depends on the relative orientation of the diffraction vector, the Burger's vector and the line vector of dislocations, and the elastic constants of the material \cite{Ungar1999a}. For polycrystalline samples, the average contrast factor of all the grains and dislocations is determined in the CMWP fitting procedure. The modified Williamson-Hall plots in Figure \ref{fig:i11}(b) are  produced following the procedures described in \cite{Ungar1996, Ungar1999a}. They demonstrate that strain anisotropy in these samples can indeed be accounted for by considering the contrast factor fitted from CMWP, as the integral peak widths are a linear function of $K^{2}C$ \cite{Borbely2022}.
				
		
		CMWP line-profile analysis fits the microstructural parameters including average crystallite size, dislocation density and the contrast factors, which describes the average character of the dislocations (`$q$ parameter') \cite{Ribarik2019, Ungar1999a}. The results of this analysis are shown in Figure \ref{fig:i11}(c)-(e). A combined procedure using successive applications of a Monte Carlo (MC) statistical algorithm and Levenberg-Marquardt nonlinear least-squares algorithm fits the peaks and physical parameters of the diffraction patterns. The error bars presented describe the range for each parameter where the corresponding fit had a resultant weighted sum of square residuals (WSSR) that fall within $p$\% confidence value in the MC procedure \cite{Tyralis2013}. At least 2000 iterations were completed during the MC fitting, and $p$ = 3.5 was used as it has been previously shown to yield a good fit, even for complicated patterns. Further details of the fitting procedure can be found in \cite{Ribarik2019, Ribarik2020}.
		
		The crystallite size fitted from CMWP for the undeformed material is $\sim$ 300 nm (Figure \ref{fig:i11}(c)), which is much smaller than that measured by EBSD (Figure \ref{fig:SEM}(a) and \ref{fig:grainsize}(a)). This is to be expected as X-rays are much more sensitive to smaller angular misorientations ($<$ 1$^{\circ}$), hence effectively mapping out regions corresponding to `subgrains' in the undeformed material \cite{Ungar2005}. For grains larger than about two microns, the size contribution to X-ray peak broadening is not possible to measure as its breadth becomes smaller than the instrument broadening contributions. Therefore, XRD is unable to accurately determine the size of `grains' in the as-received material in this study. In contrast, SEM-based methods are useful for measuring angular misorientations of 10 - 15$^{\circ}$, and on the lengthscale of microns, mostly resulting in the size measurement of `grains' (Figure \ref{fig:grainsize}(a)). 
		
		After deformation via HPT to $\gamma = 110$, the average crystallite size is 42 nm. This value is comparable to the TKD grain size measurements (Figure \ref{fig:grainsize}(b)). Irradiation does not cause any significant change in crystallite size for the undeformed material. There is a slight irradiation-induced increase in crystallite size for the HPT-deformed material. 
		
		Dislocation density increases by a factor of $\sim10$ following HPT (Figure \ref{fig:i11}(d)). Subsequent irradiation of the deformed material causes a 20\% reduction in dislocation density. In contrast, the undeformed reference material shows a slight increase in dislocation density following irradiation, likely due to irradiation-induced defects. The reduction in strain following irradiation of the deformed material can also be directly observed from the changes in peak broadening of the diffraction spectrum (Figure \ref{fig:i11}(a)) and the slope of the modified Williamson-Hall plot (Figure \ref{fig:i11}(b)). While the results of the deformed material shown in Figure \ref{fig:i11} were measured for $\gamma \sim 110$, similar results (not shown here) were also observed for $\gamma \sim 230$. This demonstrates that self-ion irradiation removes some effects of the dislocations introduced by severe plastic deformation. 
		
		The $q$ parameter describes the average character (screw vs. edge) of the dislocations present in the material. Its theoretical range depends on the elastic anisotropy ($A_{i} = 2c_{44}/(c_{11} - c_{12})$) and the ratio $c_{12}/c_{44}$) \cite{Ungar1999a}. For Eurofer-97, \textit{ab initio} calculations give $A_{i} \approx 1.5$ and $c_{12}/c_{44} \approx 1$ \cite{Li2018}. This leads to a corresponding $q$ range of between 1 (pure edge dislocations) to 2.5 (pure screw dislocations). Our fitting of the $q$ parameter suggests that the dislocations present in the undeformed material ($q \sim 2.3$) are predominantly of a screw-type character, and after HPT ($q \sim 1.4$) they are of more edge-type character (Figure \ref{fig:i11}(e)). Irradiation appears to have little effect on the $q$ parameter.

		\subsubsection{In-Situ Annealing X-ray Diffraction Measurements} \label{sec:i12}
		\begin{figure}[h!]
			\centering
			\includegraphics[width=0.95\textwidth]{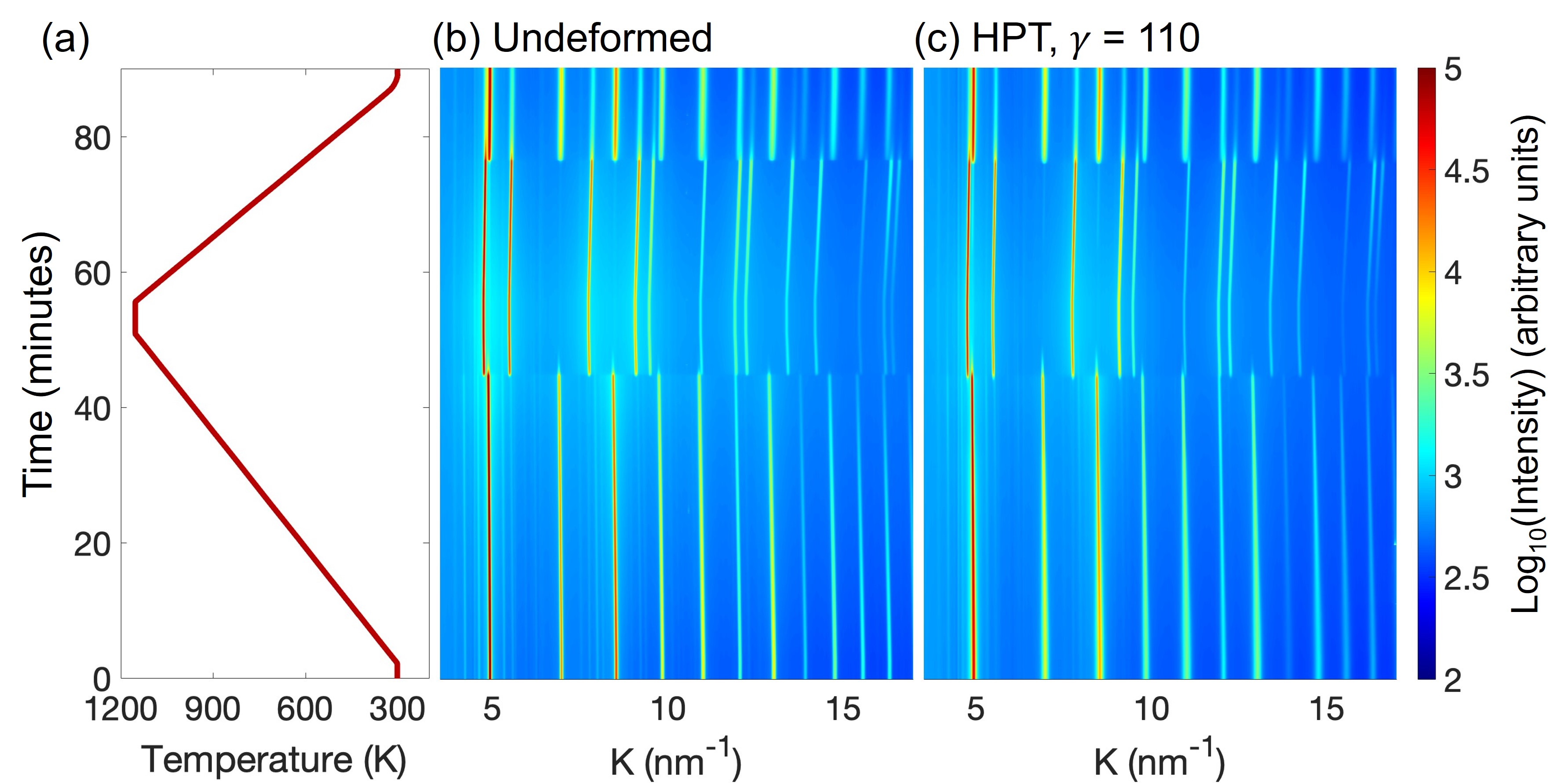}
			\caption{(a) The calibrated temperature profile of the annealing cycle during the in-situ XRD measurements. The diffraction intensity profiles as a function of $K = 2\sin\theta/\lambda$ and annealing cycle time are shown for (b) the reference undeformed sample, and (c) the HPT-deformed sample, respectively. The vertical axes representing time in (b) and (c) are aligned with (a), from which the annealing temperature can be read. }
			\label{fig:qtemp}
		\end{figure}
		
		
		In the reference undeformed sample, the peak widths did not change as a function of temperature up to 1043 K, at which point a phase transition took place (Figure \ref{fig:qtemp}(b)). The temperature at which the phase transition from ferrite/martensite (body-centred cubic/tetragonal) to austenite (face-centred cubic) take place can depend on the heating rate. In other studies of Eurofer-97 in the literature, the transition begins to take place between 1025 K - 1133 K \cite{Paul2006, Danon2002, Kumar2021}. In this study, our measurements show that all ferrite/martensite peaks disappeared at 1093 K during the heating ramp. The material contains only austenite phase during holding at 1153 K, and during the cooling stage until the material cooled to 606 K. 
		The martensite phase returned with cooling below 606 K. This is consistent with previous findings in Eurofer-97 annealing experiments \cite{Danon2002}. However, some austenite phase is still present in the material with broader diffraction peaks. Interestingly, the martensite peaks are also wider following the heating and cooling cycle than in the initial state. This could be due to residual strain in the martensite phase as it forms within the austenite phase upon cooling \cite{Hidalgo2017, Tirumalasetty2012}. There is a volume expansion associate with the formation and growth of martensite regions within the austenite grains. This will impart a hydrostatic pressure within the austenite regions, which will also in turn cause residual strains in the martensite grains \cite{Nakada2016, Villa2012}. Appendix E shows a different set of in-situ annealing XRD measurements, in which a phase transition occurred for a shorter period of time, and no austenite phase was retained following cooling. The martensite peak widths following cooling remained as narrow as right before the phase change at 973 K. This is further evidence that the post-annealing peak broadening is associated with the retained austenite phase.
		
		During heating of the HPT-deformed material, the peak widths are initially large, then gradually reduce until around 900 K (Figure \ref{fig:qtemp}(c)). Similar to the undeformed sample, a phase transition from ferrite/martensite to austenite occurs at 1043 K, then from austenite back to martensite at 606 K during the cooling stage. Both austenite and martensite phases remain in the material even at room temperature following the heating cycle. Furthermore, the peaks after cooling, for both the martensite and austenite phases, are broader than before and during the annealing cycle, respectively.

		\begin{figure}[h!]
			\centering
			\includegraphics[width=\textwidth]{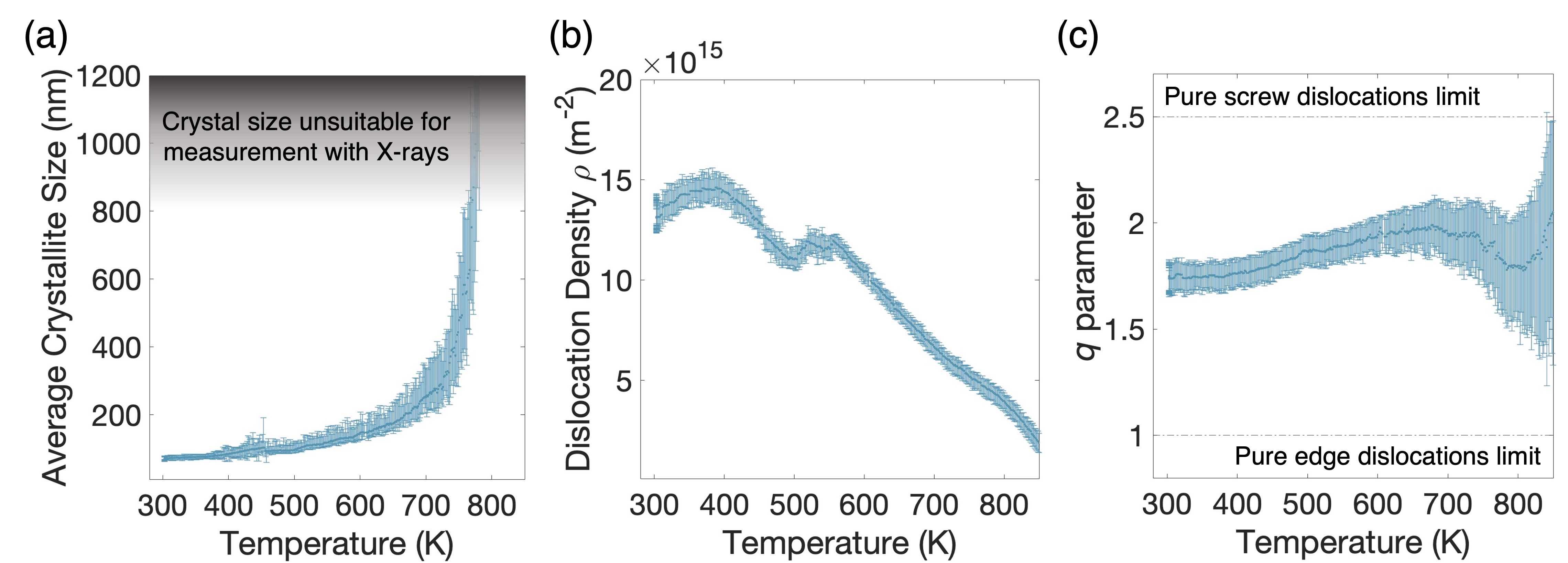}
			\caption{The microstructural parameters of the in-situ annealing diffraction measurements determined from CMWP analysis. An explanation of each parameter is included in the main text. The error bars are from the fitting algorithm discussed in Section \ref{sec:i11}.}
			\label{fig:i12}
		\end{figure}
		
		From the CMWP fitting, the area-weighted average crystallite size is unchanged between 303 K and 367 K ($\sim$ 75 nm) during the heating stage (Figure \ref{fig:i12}(a)). Further increase in temperature causes a corresponding  increase in crystallite size until $\sim$ 725 K, where the average crystallite size is around 280 nm. After this point, the crystallite size increases dramatically with temperature, suggesting that the size-broadening contribution in the diffraction patterns is approaching zero. However, we note that recrystallisation has likely not progressed far at this temperature as the XRD patterns still showed continuous Debye-Scherrer rings up until the phase transition (see supplementary files for raw patterns).
		
		There appear to be three distinct stages of evolution for dislocation density (Figure \ref{fig:i12}(b)). Between 303 K to 550 K, the dislocation density increases slightly and then decreases but the average remains within 10\% of the initial value ($\sim1.3\times10^{16}$ m$^{-2}$). Between 550 K to 800 K, $\rho$ decreases linearly to $\sim4\times10^{15}$ m$^{-2}$. Further reduction in $\rho$ above 800 K appears to be at a greater rate than in the previous stage, but it is also likely that the strain-induced peak broadening is too small to be quantitatively probed in this particular experiment.  
		
		The $q$ parameter, which is indicative of the average nature of the dislocation character, also changes during annealing (Figure \ref{fig:i12}(c)). At temperatures above 387 K, the $q$ parameter slowly increases from 1.75 to 2 at 680 K, after which the fitting from CMWP analysis appears to be less reliable. This suggests a general trend of the average dislocation character from more edge-like to more screw-like, recovering features of the undeformed material ($q\sim2.3$, Figure \ref{fig:i11}(e)). 
		
		\subsection{Material Property Changes}
		
		\begin{figure}[h!]
			\centering
			\includegraphics[width=\textwidth]{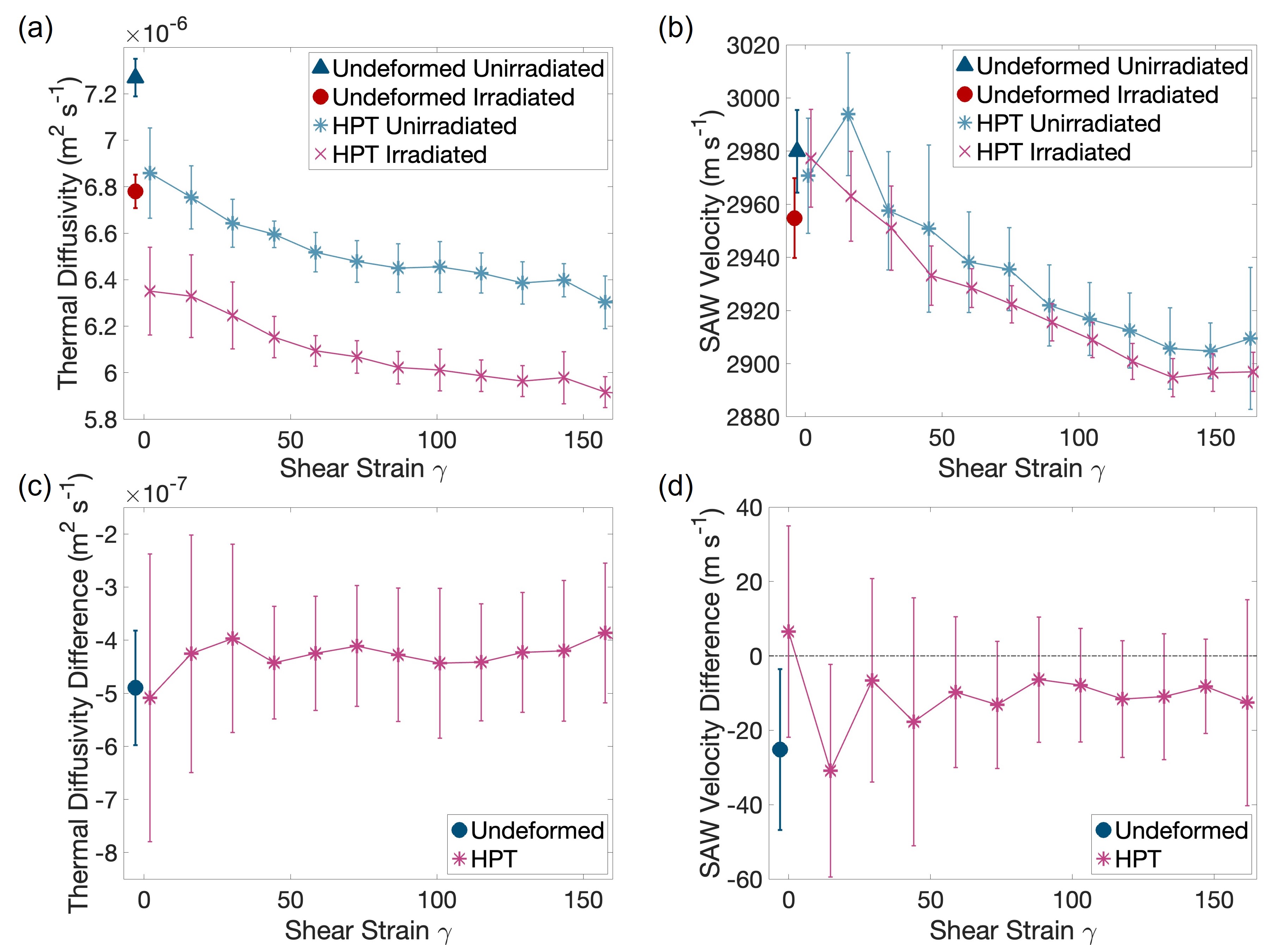}
			\caption{(a) Thermal diffusivity and (b) surface acoustic wave (SAW) velocity of Eurofer-97, comparing the reference (dark blue triangle), irradiated (red circle), deformed (light blue asterisk), and deformed then irradiated samples (pink cross). The difference in (c) thermal diffusivity and (d) SAW velocity following irradiation for the undeformed (blue circle) and deformed (pink asterisk) set of materials. The values for the deformed material are obtained by sampling points along the diameter of the specimen to measure areas of different shear strains. }
			\label{fig:tgs}
		\end{figure}
		
		Thermal diffusivity of the reference Eurofer-97 measured by TGS in this study (Figure \ref{fig:tgs}(a)) is in agreement to within 10\% of previous measurements using the laser flash technique ($8.3\times10^{-6}$ m$^{2}$s$^{-1}$) \cite{Mergia2008}. Following irradiation, the thermal diffusivity of the undeformed material decreases by 6.7\%, which is comparable to the results of FeCr \cite{Song2020} and stainless steel 316 L \cite{Graham2021} exposed to similar damage doses. 
		It is well-known that irradiation can cause a decrease in thermal diffusivity due to the introduction of defects to the crystal that act as scattering sites for electrons. Since thermal conductivity in metals is predominately electron-mediated, an increase in electron scattering rate results in a reduction of thermal diffusivity \cite{Wylie2022}.
		
		Plastic deformation via HPT also has the effect of reducing the thermal diffusivity of Eurofer-97 (Figure \ref{fig:tgs}(a)). It is interesting to note that at nominally zero shear strain (at the centre of the HPT sample), the thermal diffusivity is not the same as the reference as-received material. This difference shows that even though the centre experiences nominally only compression, which does not cause significant grain refinement, there is still a significant reduction in thermal diffusivity. This suggests that there is an appreciable increase in dislocation density in the sample centre during the initial compression. 
		
		Thermal diffusivity decreases with increasing shear strain (Figure \ref{fig:tgs}(a)). At $\gamma = 160$, the maximum strain level measured on the HPT samples due to the sample geometry, there is a decrease of 12\% from the reference Eurofer-97 value. HPT deformation increases the dislocation density in the material and increases the ratio of grain boundaries to perfect crystal. Both these factors lower the mobility of charge carriers in the material and thus cause a reduction in thermal diffusivity \cite{Badry2020, Dong2014}. The rate of thermal diffusivity decrease appears to reduce with increasing shear strain, particularly at $\gamma > 70$. 
		
		Irradiation of the deformed material causes a further reduction in thermal diffusivity in addition to the decrease associated with the deformation. At $\gamma = 160$ in the irradiated sample, there is a 19\% reduction in thermal diffusivity compared to the reference unirradiated Eurofer-97 sample. 
		
		Following irradiation, the material that has undergone prior HPT shows a smaller reduction in thermal diffusivity compared to the undeformed material (Figure \ref{fig:tgs}(c)). This could be due to the irradiation-induced defects being absorbed by the greater proportion of grain boundaries present in the HPT material \cite{Enikeev2019}, therefore causing less change to the thermal properties. There could also be competing effects between the introduction of irradiation-induced defects and the irradiation process itself causing simultaneous annealing of the pre-existing dislocations from HPT (Section \ref{sec:i11}). As these two processes act in opposition to each other, the net change to thermal diffusivity after deformation and irradiation is smaller than for irradiation alone. However, we also note that the irradiation-induced difference in thermal diffusivity between the undeformed and deformed cases is small, and could be within measurement error. 
		
		Surface acoustic wave (SAW) velocity is related to the elastic constants of the material. However, since Fe and Fe-based alloys have strong elastic anisotropy, the relationship between SAW velocity and the elastic constants is dependent on the grain orientation or local texture \cite{Hofmann2019}. Since our grains are many times smaller than the measurement spot even in the undeformed case, we cannot quantitatively measure the changes to the elastic constants in this study. However, changes to the SAW velocities were still measured, indicating that the material stiffness has been altered by HPT and irradiation (Figure \ref{fig:tgs}(b)).
		
		A 1\% reduction in SAW velocity due to irradiation in the undeformed sample is observed. This could be due to the creation of point defects from irradiation. Previous studies with He-ion irradiation in tungsten  have shown a $\sim$ 2\% reduction in SAW velocity at 0.2 - 0.3 dpa \cite{Hofmann2015, Duncan2016}. However, studies in copper and nickel (different crystal structures to Eurofer-97) have shown an increase in SAW velocity for irradiation up to 5 dpa, which the authors attributed to the interaction of defect clusters with dislocations causing a pinning effect \cite{Dennett2018, Dennett2019}.
		
		Deformation causes a significant reduction in SAW velocity (Figure \ref{fig:tgs}(b)). Note that due to the measurement geometry, the SAW velocities reported here are in the radial direction $\boldsymbol{r}$ (perpendicular to the shear direction $\boldsymbol{\theta}$). At $\gamma = 160$, the SAW velocity has decreased by 2.5\% compared to the undeformed material. There also does not appear to be any saturation effect until $\gamma = 140$. This is a significantly greater level of strain than required for the saturation of thermal diffusivity changes (Figure \ref{fig:tgs}(a)) and hardness  \cite{Strangward-Pryce2023}, both of which approached saturation at $\gamma > 70$. Grain size and GND density, as observed by TKD, have both saturated by $\gamma = 110$ (Figure \ref{fig:grainsize} and \ref{fig:gnd}), which is also lower than the shear strain threshold for SAW velocity change saturation.
		
		Irradiation of the deformed material causes a further reduction in SAW velocity, however this could be within measurement error. The change to SAW velocity due to irradiation is less for the materials that have undergone prior HPT deformation compared to the undeformed specimens (Figure \ref{fig:tgs}(d)). This could again be due to a combination of the grain boundaries absorbing irradiation defects and annealing of dislocations caused by HPT during the process of irradiation.

	\section{Discussion}
		\subsection{Effect of HPT}
		HPT causes significant grain refinement in Eurofer-97. From SEM-based analysis, the area-weighted average grain size decreases from 5.26 \textmu m to 146 nm, which is a reduction by a factor of $\sim$ 30 (Figure \ref{fig:grainsize}). From CMWP analysis of the grazing-incidence XRD data, the area-weighted average crystallite size decreases from 295 nm to 42 nm, which is a reduction by a factor of $\sim$ 7 (Figure \ref{fig:i11}(c)). From CMWP analysis of the transmission XRD measurements, the grain size of the HPT-deformed sample before annealing is 75 nm (Figure \ref{fig:i12}(a)). Though the grain size reduction ratio from HPT is different between electron microscopy and XRD, all measurements of the grain sizes following deformation are in good agreement. This has been observed from other studies of deformed materials that compare electron microscopy techniques with XRD \cite{Zhilyaev2003, Gubicza2004}.
		
		The discrepancies between grain size measurements of the undeformed material from EBSD and XRD can be attributed to their respective sensitivity to misorientations in the lattice (Section \ref{sec:i11}). As the XRD coherent scattering domains are small, the crystallite size measurements are more indicative of `subgrain' size. On the other hand, for EBSD, the misorientation threshold used 10$^{\circ}$, which allows the measurement of `grain' size.  However, the agreement between TKD and XRD measurements for the grain and crystallite size of the HPT-deformed material suggests that HPT causes the formation of high-angle grain boundaries. The distinction between `grains' and `subgrains' ceases to exist following grain refinement. This is also supported by the greater distribution of grain boundaries with $>$30$^{\circ}$ misorientation of the HPT material compared to the undeformed (Figure \ref{fig:texture}(a)-(b)).
		
		The GND density measured by EBSD/TKD for the undeformed and HPT samples are respectively $9.5\times10^{14}$ m$^{-2}$ and $7.5\times10^{15}$ m$^{-2}$ (Figure \ref{fig:gnd}). In comparison, the total dislocation density values measured by grazing-incidence XRD is $2.5\times 10^{14}$ m$^{-2}$ and $3.8\times10^{15}$ m$^{-2}$, respectively for the undeformed and HPT-deformed samples (Figure \ref{fig:i11}(d)). From transmission XRD, the dislocation density of the deformed material is $1.3\times10^{16}$ m$^{-2}$ (Figure \ref{fig:i12}(b)). The factor of dislocation density increase from HPT deformation is $\sim$ 8 from EBSD/TKD, and $\sim$ 15 measured from XRD. These values are in good agreement despite the fact that EBSD/TKD only probes a thin layer near the sample surface and only GNDs. In comparison, XRD measurements average over a much larger sample volume and mainly measure statistically stored dislocations (SSDs) \cite{Gallet2023}. The comparison between TKD and XRD measurements of dislocation density in the HPT-deformed material suggests that the density of GNDs is higher than the density of SSDs. This can be rationalised by considering that in severely deformed materials, dislocations are mainly located in the vicinity of grain boundaries (also seen in Figure \ref{fig:gnd}(b)) and in the form of GNDs \cite{Zhang2020}. 
		
		For characterisation with EBSD/TKD, only the indexed points contribute to the GND density calculation, which means that sampling is skewed towards the regions further away from the grain boundaries. In contrast, XRD probes the full sample volume, including regions of strong lattice distortions close to grain boundaries, which are not indexed with EBSD/TKD. The detection of a greater increase in dislocation density following HPT from XRD measurements compared to EBSD/TKD supports the theory that large amounts of dislocations are stored in the grain boundaries rather than the grain interiors \cite{Valiev2000, Valiev1993}. It is important to note that for EBSD/TKD, the GND density is related to the degree of misorientation between adjacent points, and hence dependent on the step size used in the measurement. A smaller step size leads to more measurement noise \cite{Jiang2013}. The step size used for the EBSD of the undeformed sample is 25 times larger than the step size used for TKD in the deformed sample, due to the measurement constraints and the vastly different grain sizes in each sample. Therefore, the GND density in the HPT-deformed samples could be slightly overestimated.		
		
		HPT processing also causes dislocation character to change from more screw-like to more edge-like (Figure \ref{fig:i11}(e)). At room temperature in coarse-grained BCC metals, the dislocation population is predominantly of screw type, which is the case for the undeformed Eurofer-97. This is due to the high Peierl's barrier and low mobility of screw dislocations \cite{Vitek2011}, leading to a greater retention than edge dislocations \cite{Duesbery1998}. Interestingly, it has been observed in other nanocrystalline BCC metals produced by HPT including tungsten \cite{Wei2006}, molybdenum \cite{Cheng2013, Cheng2013a} and tantalum \cite{Wei2011}, there is an increased population of dislocations with edge character compared to the coarse-grain material. There are several explanations proposed to explain these observations. One of which hypothesises that the smaller grain sizes, and hence a greater proportion of grain boundaries, means that fewer dislocations are retained in the crystal as they are annihilated at the grain boundaries \cite{Cui2016}. This reduces the significance of the difference in mobility between edge and screw dislocations, leading to more comparable populations of each type. Furthermore, HPT deformation causes a large increase in dislocation density (forest dislocations), which in turn impedes the motion of glide dislocations \cite{Hull2011}. The mobility of glide dislocations becomes increasingly governed by the forest dislocation density. This again makes the difference in mobility between screw and edge dislocations more insignificant. The other hypothesis is that hydrostatic compression increases the mobility of screw dislocations \cite{Cui2016, Huang2011}. This is because higher stress reduces the activation volume for thermally-activated kink-pair formation. This in turn reduces the thermal energy barrier for dislocation motion and assists the movement of screw dislocations at low temperatures. An increase in screw dislocation mobility would lead to a more comparable probability of retention between screw and edge dislocations. Further studies are needed to determine which mechanism, or combination of them, is at play here.  
		
		
		\subsection{Thermal Stability of HPT Eurofer-97}
		Considering the changes in crystallite size and dislocation density, the HPT-processed material first undergoes a recovery process between 450 K to 800 K (Figure \ref{fig:i12}). During the start of the recovery stage (450 K to 550 K), dislocation density is not expected to change significantly. The growth in crystallite size in this temperature range can be attributed to the rearrangement of dislocations within grains, causing subgrain coarsening whilst total dislocation density remains unchanged \cite{Raabe2014, Humphreys1996}. Further annealing causes both the growth of these subgrains as well as the annihilation of dislocations, possibly at grain boundaries \cite{Hao2014}, reducing the overall dislocation density as crystallite sizes increase. At $\sim$ 800 K, the onset of recrystallisation causes rapid grain growth and a reduction in dislocation density. Annealing also causes the average character of dislocations in the HPT Eurofer-97 material to revert from more edge-like to more screw-like, partially recovering the features of a coarse-grain BCC material \cite{Vitek2011, Duesbery1998}.
		
		Other thermal stability studies of HPT-processed steel have been reported in the literature. For stainless steel 316L with $\gamma$ up to 360, the dislocation density and crystallite size prior to annealing were comparable to this study, and these quantities stayed stable up to 560 K after which dislocation density reduced significantly \cite{El-Tahawy2017}. It is also interesting to note, that in that study, the steel structure was initially austenitic and transformed to martensitic during the process of HPT. During annealing, it slowly transformed back to austenitic, beginning at 650 K and up to 93\% transformed following annealing at 1000 K. A study with differential scanning calorimetry of a Fe-10Ni-7Mn martensitic steel processed by HPT up to $\gamma = 785$ showed that dislocation annihilation with vacancies began around 504 K \cite{JavadzadehKalahroudi2019}. This is similar to the temperature at which dislocation density started to reduce substantially in this study. 
		
		\subsection{Effect of Irradiation}
		Ion irradiation to $\sim0.1$ dpa did not cause any microstructural changes that were observable by TEM or TKD. However, clear changes were measured with XRD and TGS. The reduction in dislocation density and increase in crystallite size following irradiation in the HPT-processed samples (Figure \ref{fig:i11}(c)-(d)) suggest an irradiation-induced-annealing type process. Similar effects have been observed following irradiation of HPT-processed austenitic stainless steel 316 \cite{Radiguet2008}, martensitic Fe-Cr-W steel \cite{Mazilkin2020}, EK-181 \cite{Aydogan2017}, and T91 steel \cite{Gigax2017}. 
		However, these other studies typically examined much higher doses ($>$ 1 dpa) and elevated temperatures ($>$ 673 K). 
		Observations of crystallite growth and dislocation density reduction in our study suggest that the movement of dislocations and subgrain coarsening aided by irradiation can happen at much lower irradiation dose and temperature.
		
		A previous nanoindentation study of this same set of materials found that HPT-processing hardened the material significantly, approaching saturation beyond $\gamma = 70$ \cite{Strangward-Pryce2023}. Subsequent irradiation to $\sim$ 0.01 dpa caused slight softening, but further hardening is observed following irradiation to $\sim$ 0.1 dpa. The magnitude of hardening changes due to irradiation is much smaller than from HPT deformation. 
		
		In light of the previous hardness results \cite{Strangward-Pryce2023} and the findings from the present study, we can draw a combined picture of the effect of HPT and irradiation. 
		HPT introduces grain refinement and a large population of dislocations due to plastic deformation. Subsequent irradiation introduces point defects and dislocation loops into the material. This is supported by the fact that the dipole character parameter fitted from CMWP suggests that the dislocations are arranged in a stronger dipole formation following irradiation, a possible indication of dislocation loops (Appendix D). However, from grazing-incidence XRD measurements we see a reduction of total dislocation density following irradiation in the HPT material compared to the unirradiated HPT material (Figure \ref{fig:i11}). This indicates a relaxation or annihilation of dislocations due to irradiation. 
		
		HPT deformation and irradiation both individually cause an increase in hardness, and a decrease in SAW velocity and thermal diffusivity of Eurofer-97. The effect of HPT alone causes more changes to the material properties, and total dislocation density, than irradiation alone. The effects of HPT deformation and irradiation to $\sim$ 0.1 dpa on the material properties are additive, whereas for dislocation density, subsequent irradiation removed some dislocations from prior deformation. Thus, we can conclude that changes to material properties are not simply a function of the total dislocation density but on their type and lengthscale. The deformation-induced extended dislocations annihilated in the HPT material during irradiation at $\sim$ 0.1 dpa do not contribute as much change in material properties as the small defects introduced by irradiation. However, the irradiation softening at $\sim$ 0.01 dpa \cite{Strangward-Pryce2023} could indicate that the initial removal of dislocations from HPT deformation is significant for material hardening. On the other hand, while the more dispersed but smaller irradiation-induced point defects and dislocation loops contribute less to XRD peak broadening, they still have a significant effect on material properties. This is particularly noticeable for changes to thermal diffusivity. The additional reduction of thermal diffusivity following irradiation of the HPT-deformed material is comparable to the reduction from solely the HPT process. The effect of small point defects, such as vacancies, on thermal diffusivity in other irradiated materials has been previously investigated \cite{Reza2020}.
		
		
		


	\section{Summary and Conclusion}
		
		In this study, grain refinement in Eurofer-97 has been successfully achieved via HPT at room temperature. The thermal stability and irradiation response of the resultant material have been characterised. The key findings are as follows:
		\begin{itemize}
			\item Grain refinement via the creation of high-angle grain boundaries was achieved with HPT. Grains smaller than 100 nm were created and the average grain size reduced from 5.26 \textmu m to 146 nm. 
			\item The average dislocation density in the material increased by over an order of magnitude following HPT, up to $10^{16}$ m$^{-2}$. The majority of these dislocations are located at or near grain boundaries.
			\item Grain refinement from HPT increased the proportion of dislocations with edge-like character. This is consistent with observations of other nanocrystalline BCC materials in literature. 
			\item Saturation of grain size refinement and dislocation density increase was achieved by $\gamma = 110$. Thermal diffusivity and SAW velocity changes saturated at $\gamma = 70$ and $\gamma = 140$, respectively. The saturation of microstructure and material properties at different shear strains indicates that they are governed by different mechanisms.
			\item Recovery was observed during annealing between 450 K to 800 K, with subgrain coarsening from the rearrangement and subsequent annihilation of dislocations. Recrystallisation and rapid grain growth begin around 800 K.
			\item Irradiation of the undeformed Eurofer-97 material caused an increase in dislocation density. However, for the HPT-deformed material, irradiation caused a reduction in dislocation density, suggesting an irradiation-induced annealing process.
			\item HPT processing prior to irradiation did not prevent additional irradiation-induced reductions in thermal diffusivity and SAW velocity, which also occurred in the undeformed case. The irradiation-induced changes are slightly smaller in magnitude in the HPT materials compared to the undeformed but the effect is not significant.			
		\end{itemize}		
		
		By considering the microstructure and material properties of Eurofer-97, a multi-faceted view of the effect of SPD, and its subsequent impact on thermal stability and radiation resistance for ferritic/martensitic steels, emerges.
		
	\section*{Declarations}
		\subsection*{Funding}
		The  authors  acknowledge  use  of  characterisation  facilities  within  the  David  Cockayne  Centre for  Electron  Microscopy,  Department  of  Materials,  University  of  Oxford,  alongside  financial support  provided  by  the  Henry  Royce  Institute (grant  ref EP/R010145/1). KS acknowledges funding from the General Sir John Monash Foundation and the University of Oxford Department of Engineering Science. FH, DY and AR acknowledge funding from the European Research Council (ERC) under the European Union’s Horizon 2020 research and innovation programme (grant agreement No. 714697). DEJA acknowledges funding from EPSRC grant EP/P001645/1. This material is based upon work done at Brookhaven National Laboratory, supported by the U.S. Department of Energy (DOE), Ofﬁce of Basic Energy Sciences, under Contract No. DE SC0012704.
		
		\subsection*{Conflicts of interest}
		The authors have no relevant financial or non-financial interests to disclose.
		
		\subsection*{Data and code availability}
		All data, raw and processed, as well as the processing and plotting scripts are available at: \textit{A link will be provided after the review process and before publication.}
		
	\section*{Acknowledgements}
		The authors are grateful for the assistance provided by DLS beamline scientists Stephen Thompson and Eamonn Connolly (I11), Robert Atwood and Stefan Michalik (I12). The authors are also grateful for the assistance from Andy Bateman and Simon Hills (Department of Engineering Science, University of Oxford) for sample cutting and preparation.
	
	\newpage
	\renewcommand{\thefigure}{A-\arabic{figure}}
	\setcounter{figure}{0}
	\section*{Appendix A}
		During the in-situ annealing experiment at DLS I12 beamline, the temperature was initially recorded as the temperature read-out ($T_{L}$) from the Linkam heating stage device. However, due to imperfections in the thermal contact between the heater and the sample, this read-out value may not be exactly the same as the actual temperature of the sample ($T_{s}$). In order to find the relationship between $T_{L}$ and $T_{s}$, the lattice parameter of the sample (calculated from the measured diffraction data) and the coefficient of thermal expansion were used. For simplicity, this relationship was approximated as linear and the calculations only consider the patterns up to the point of ferrite/martensite-to-austenite phase transition.
		
		\begin{figure}[h!]
			\centering
			\includegraphics[width=0.7\textwidth]{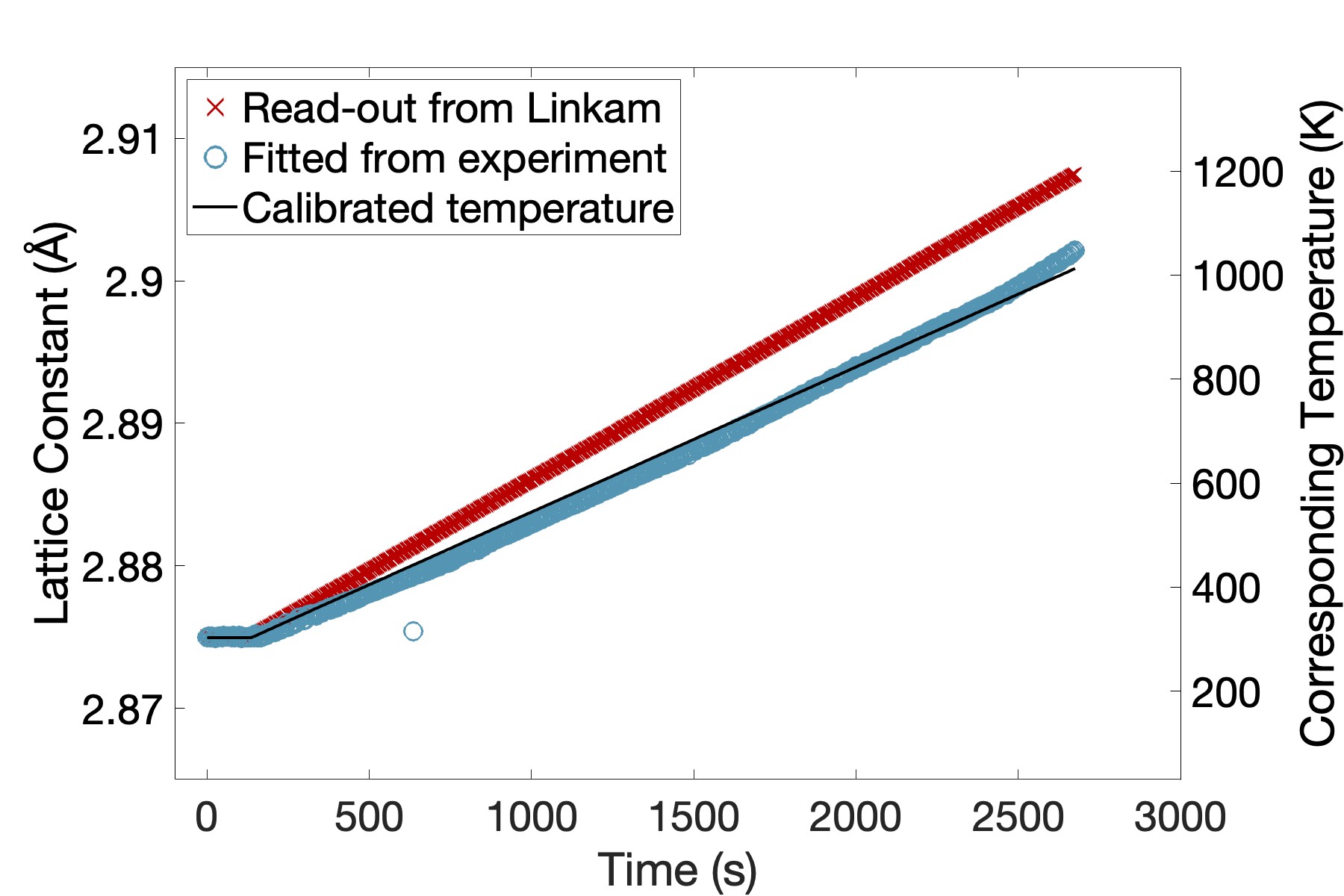}
			\caption{The lattice constant and corresponding temperature calculated and measured from the experimental diffraction patterns, and the Linkam temperature read-out. The method for obtaining the calibrated temperature curve is explained in the accompanying text.}
			\label{fig:temp}
		\end{figure}
		
		Consider the undeformed Eurofer-97 sample as an example. Figure \ref{fig:temp}, considering only the left-hand-side axis for now, shows the lattice constant fitted from the experimental diffraction patterns ($a_{s}$). It also shows the lattice constant predicted from the Linkam temperature read-out ($a_{L}$) as a function of time. $\textsl{a}_{L}$ at a certain temperature $T_{L}$ is given by: 
		\[ a_{L (T = T_{L})} = a_{s (T=303 K)} + \upalpha \times a_{s (T=303)} \times (T_{L} - 303) \]
		
		where $\upalpha = 12.74 \times 10^{-6}$ is the average coefficient of thermal expansion of Eurofer-97 between 293 K to 1073 K \cite{Rieth2003}. It is assumed at 303 K, the start of the experiment, that $T_{L} = T_{s} = 303 \text{ K}$ and thus $a_{L} = a_{s} = 2.875$ \AA $ $ (fitted from experiment).
		
		Since the lattice constant is linearly dependent on temperature, the value of $T_{s}$ is at a given time can be calculated from the lattice constant by replacing $a_{L}$ and $T_{L}$ of the above equation and rearranging:
		\[T_{s} = \frac{a_{s (T = T_{s})} - a_{s (T = 303 K)}}{\upalpha \times a_{s (T = 303 K)}} + 303 \]
		
		which gives the corresponding right-hand-side y-axis of Figure \ref{fig:temp}.
		
		A linear fit was obtained for $T_{s}$ and $T_{L}$. Then a proportionality constant was obtained, assuming the linear fits for $T_{s}$ and $T_{L}$ intersected at 303 K, to correlate $T_{s}$ and $T_{L}$ (as shown in the black line of Figure \ref{fig:temp}).

	\newpage
	\renewcommand{\thefigure}{B-\arabic{figure}}
	\setcounter{figure}{0}
	\section*{Appendix B}
\begin{figure}[h!]
			\centering
			\includegraphics[width=0.85\textwidth]{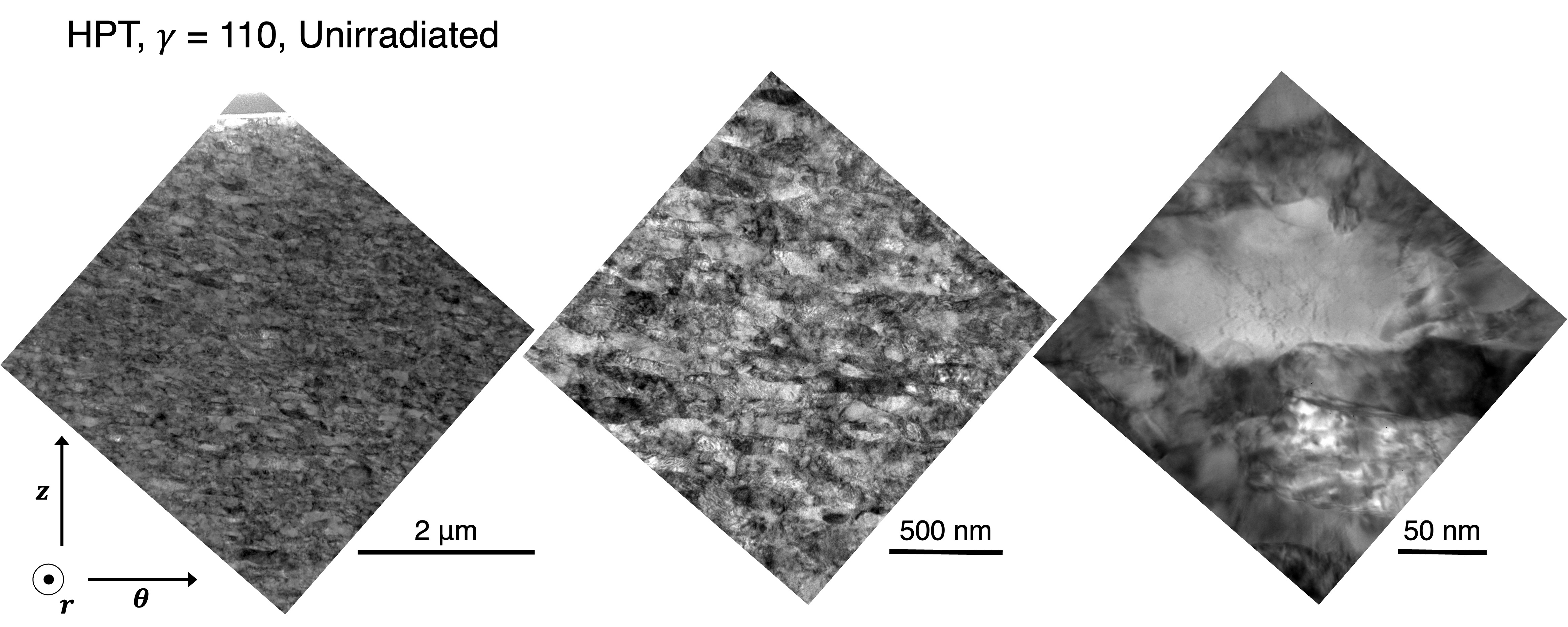}
			\includegraphics[width=0.85\textwidth]{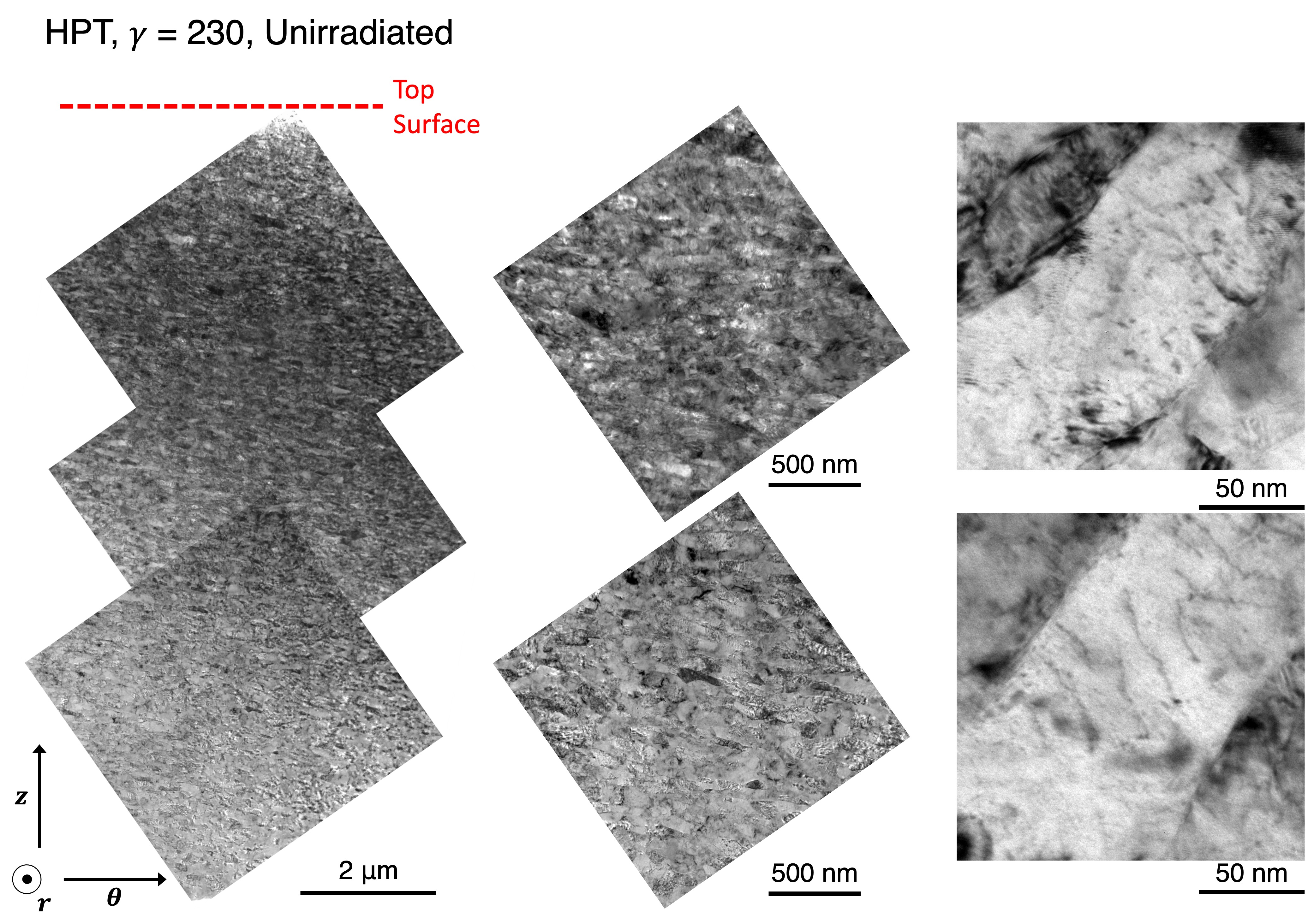}
			\caption{TEM images of the HPT sample at $\gamma = 110$ and $\gamma = 230$ with no irradiation at different levels of magnification. Dislocation lines are evident within certain grains. There is no significant depth dependence of microstructural features.}
			\label{fig:TEM1}
		\end{figure}

\begin{figure}[h!]
			\centering
			\includegraphics[width=0.85\textwidth]{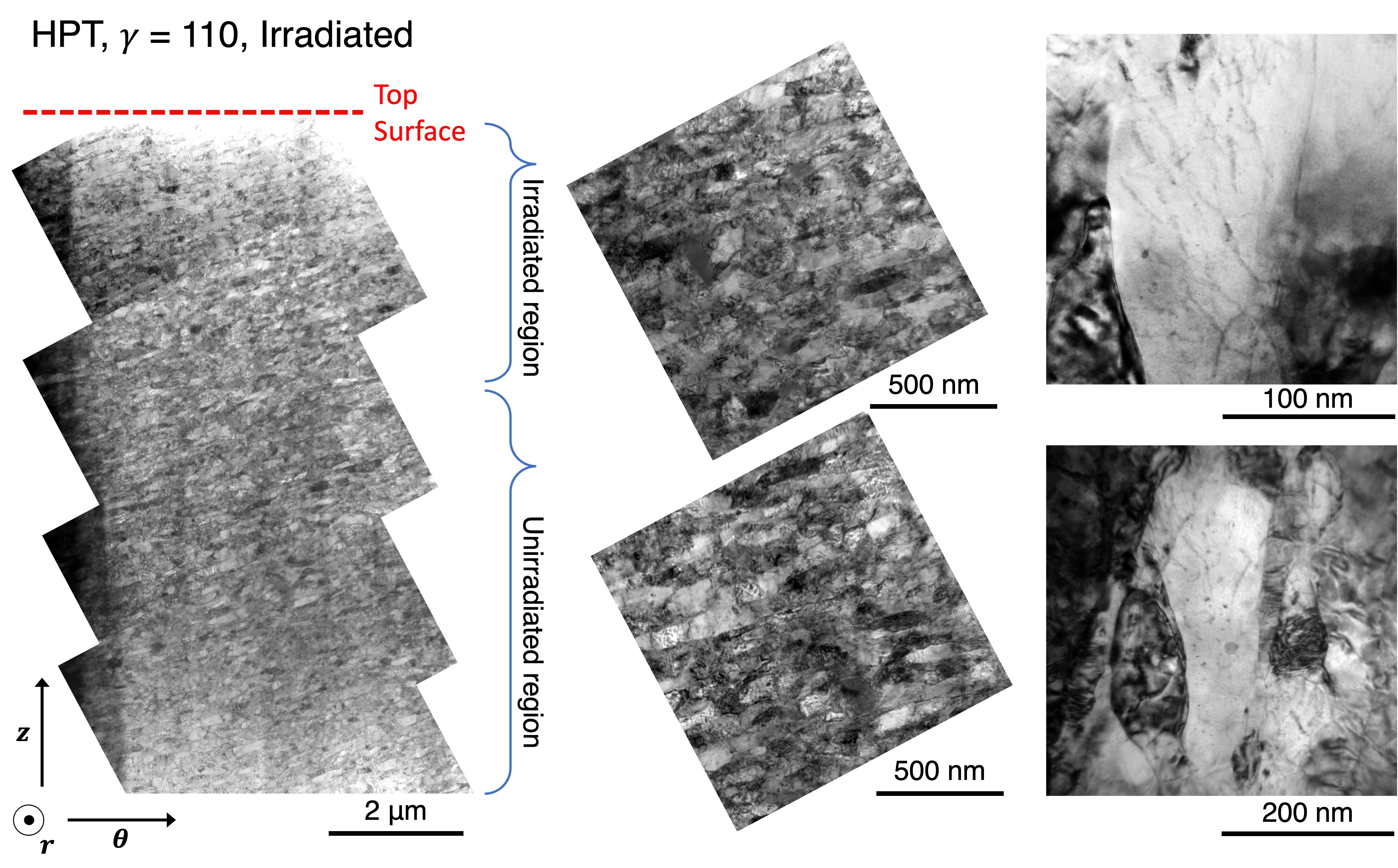}
			\caption{TEM images of the HPT sample at $\gamma = 110$ with the irradiated region indicated, and imaged at different levels of magnification. Dislocation lines are evident within certain grains. There is no depth dependence of features and irradiation defects could not be unambiguously identified in this study due to the complexity of the microstructure.}
			\label{fig:TEM2}
		\end{figure}

		The full comparison of TEM micrographs taken for each lift-out examined is included in Figures \ref{fig:TEM1} and \ref{fig:TEM2}. Low-magnification images have been rotated such that they are aligned with the coordinates indicated.
	
	\renewcommand{\thefigure}{C-\arabic{figure}}
	\setcounter{figure}{0}
	\section*{Appendix C}
\begin{figure}[h!]
			\centering
			\includegraphics[width=0.95\textwidth]{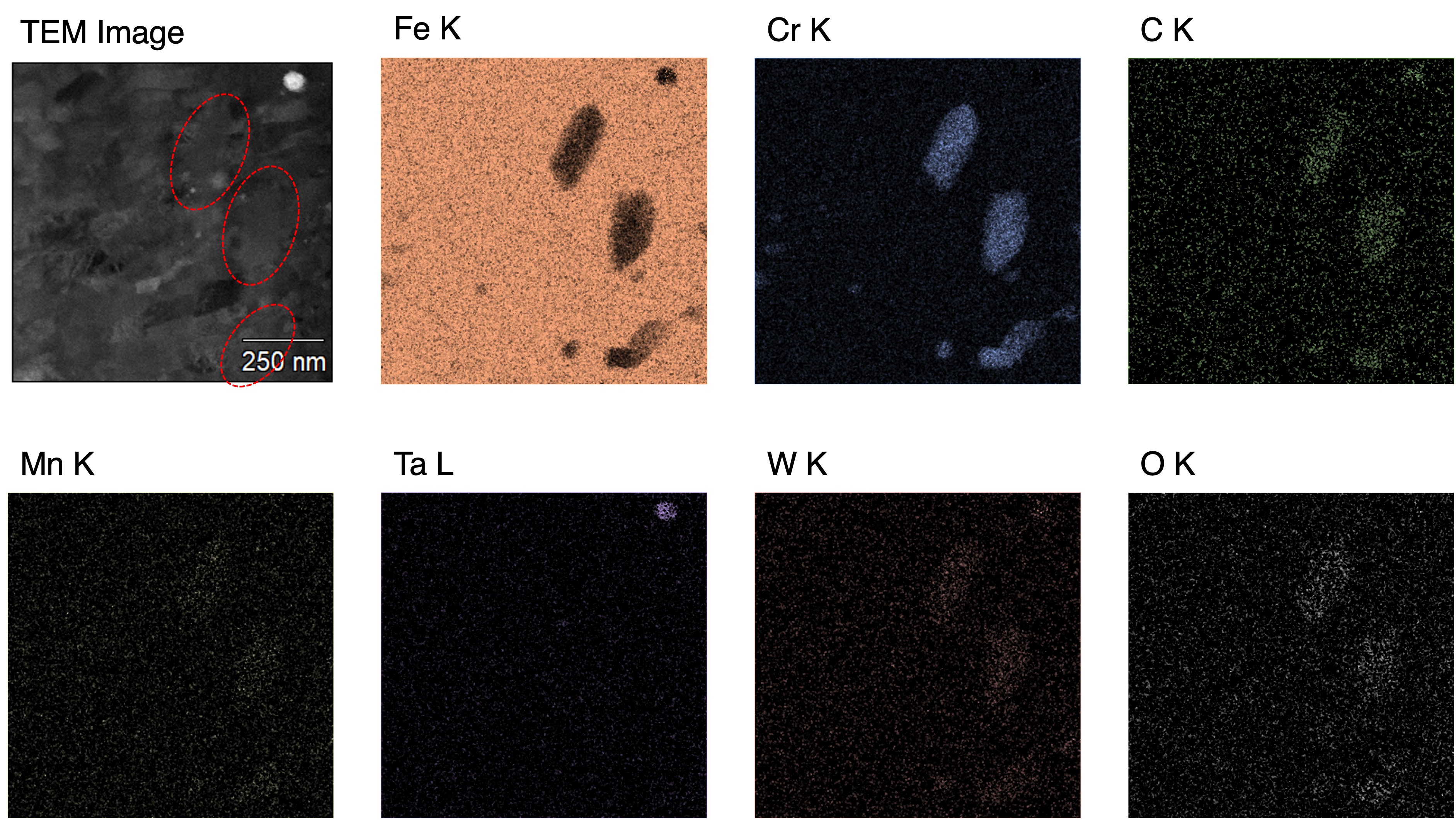}
			\caption{EDX maps of the carbide regions (circled in red) found between individual grains after HPT.}
			\label{fig:edx}
		\end{figure}

		Second-phase regions were observed in TEM and TKD as discussed in the main text. Energy dispersive X-ray spectroscopy (EDX) was performed on a TEM lift-out of the deformed ($\gamma = 230$) and irradiated sample (Figure \ref{fig:edx}). Those regions are found to be enriched with Cr (up to 30\%), W (up to 10\%), Mn and C. This is consistent with the carbides found in Eurofer-97 by another study \cite{Arredondo2020}.
		
	\newpage
	\renewcommand{\thefigure}{D-\arabic{figure}}
	\setcounter{figure}{0}
	\section*{Appendix D}
		\begin{figure}[h!]
			\centering
			\includegraphics[width=0.6\textwidth]{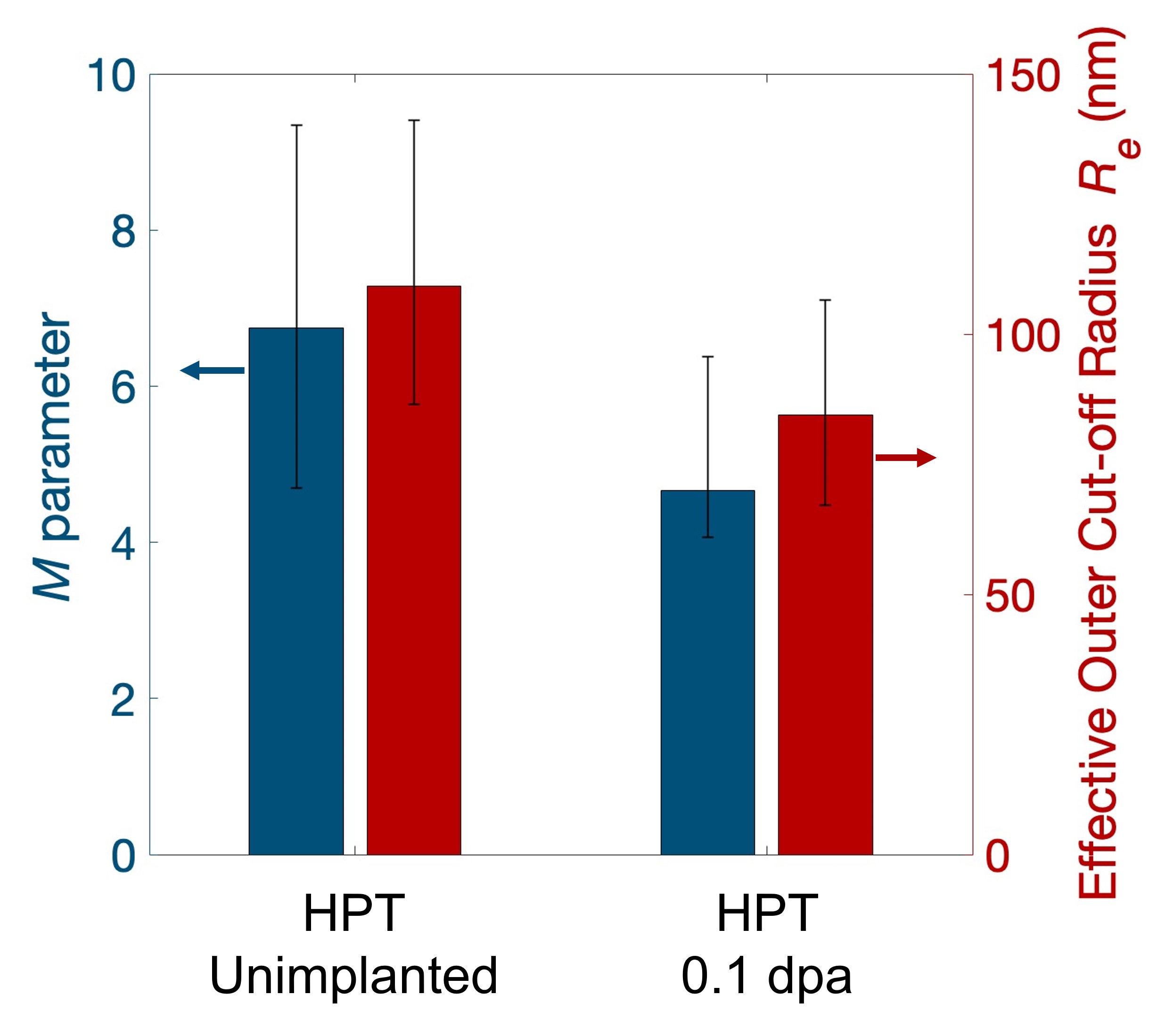}
			\caption{The values of the $M$ parameter (blue bars) and the effective outer cut-off radius of the dislocations (red bars, $R_{e}$) fitted from CMWP analysis.}
			\label{fig:MR}
		\end{figure}
		
		The effective outer cut-off radius of the dislocations ($R_{e}$) gives an indication of the range of elastic strain from the dislocations. The `randomness' of the dislocation arrangement can be described by cylinders of radius $R_{e}$ that contain equal numbers of screw dislocations of positive and negative sign, and these cylinders fill the whole crystal \cite{Wilkens1970}. Since $R_{e}$ actually depends on the dislocation density $\rho$, the dimensionless parameter $M$ is defined to give an indication of the dipole characteristic of the dislocation. It is calculated by $M = R_{e}\sqrt{\rho}$. When $M \gg 1$, the dislocations are distributed randomly and in an uncorrelated arrangement. Smaller $M$, on the other hand, indicates that dislocations of opposite sign are positioned close to each other \cite{Ribarik2020}.
		
		From CMWP fitting, reductions in the $M$ and $R_{e}$ values is seen following irradiation in the HPT-deformed sample (Figure \ref{fig:MR}). This suggests that the strain fields are becoming shorter range and possibly indicates the formation of dislocation loops \cite{Ungar2021}. To characterise the exact density of dislocation loops and line dislocations separately, the contrast factor $C$ of each type needs to be known. This has been calculated for hexagonal materials \cite{Balogh2016} but remains to be investigated for cubic materials.
	
	\renewcommand{\thefigure}{E-\arabic{figure}}
	\setcounter{figure}{0}
	\section*{Appendix E}
		The purpose of this section is to demonstrate that the peak broadening in the martensite and austenite phases post-annealing is due to the presence of the retained austenite phase. We will re-present the data from Figure \ref{fig:qtemp} here, with more analysis, for a clearer comparison with another set of in-situ annealing XRD data.
		
		\begin{figure}[h!]
			\centering
			\includegraphics[width=0.9\textwidth]{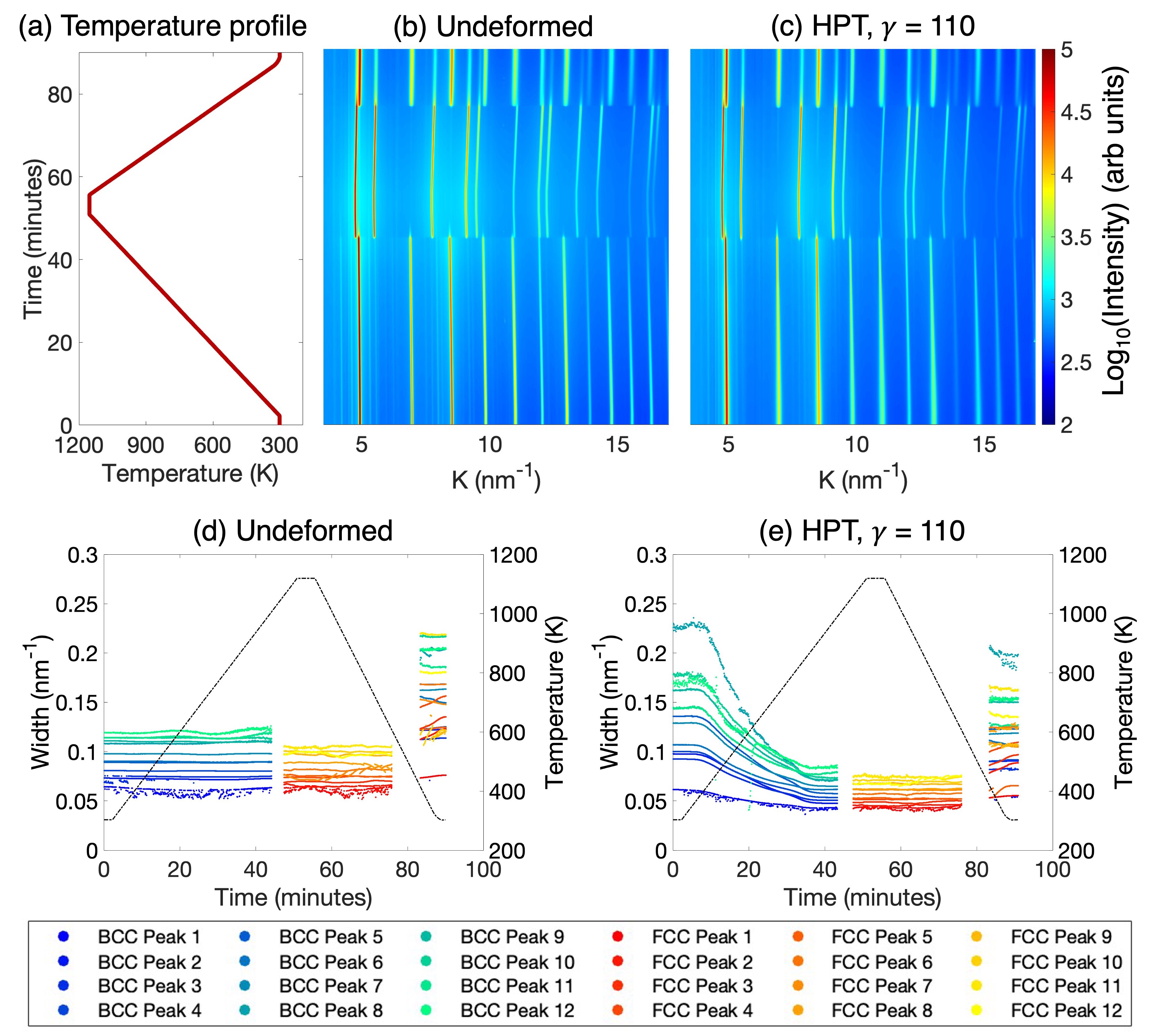}
			\caption{a) The temperature ramp and raw time vs $K = 2\sin\theta/\lambda$ plots of the b) undeformed and c) HPT sample (as discussed in Section \ref{sec:i12} of the main text). The peak widths (FWHM) of the d) undeformed and e) HPT sample as a function of time during the annealing cycle. The calibrated temperature profile is shown as the black dashed line. The ferrite/martensite phase was fitted to a body-centred cubic (BCC) structure, and the austenite was fitted to a face-centred cubic (FCC) structure. }
			\label{fig:fastramp}
		\end{figure}
		
		The fitted peak widths as a function of time is shown alongside the raw data in Figure \ref{fig:fastramp}. Key features of the raw data was discussed in Section \ref{sec:i12} of the main text. The peak temperature reached for this set of annealed samples is 1153 K. The phase transition (ferrite/martensite to austenite) occurred from around 1025 K onwards. The peak widths of both samples just before the onset of the phase transition are similar. After the annealing and cooling cycle, some austenite phase remained in the sample as evidenced in the diffraction pattern. The peak widths of both the martensite and austenite phases showed significant broadening after the martensite phase reappeared around 606 K during the subsequent cooling stage (Figure \ref{fig:fastramp}(d) and (e)). The peak width of the sample after annealing is similar in both the undeformed and HPT samples. 
		
		\begin{figure}[h!]
			\centering
			\includegraphics[width=0.9\textwidth]{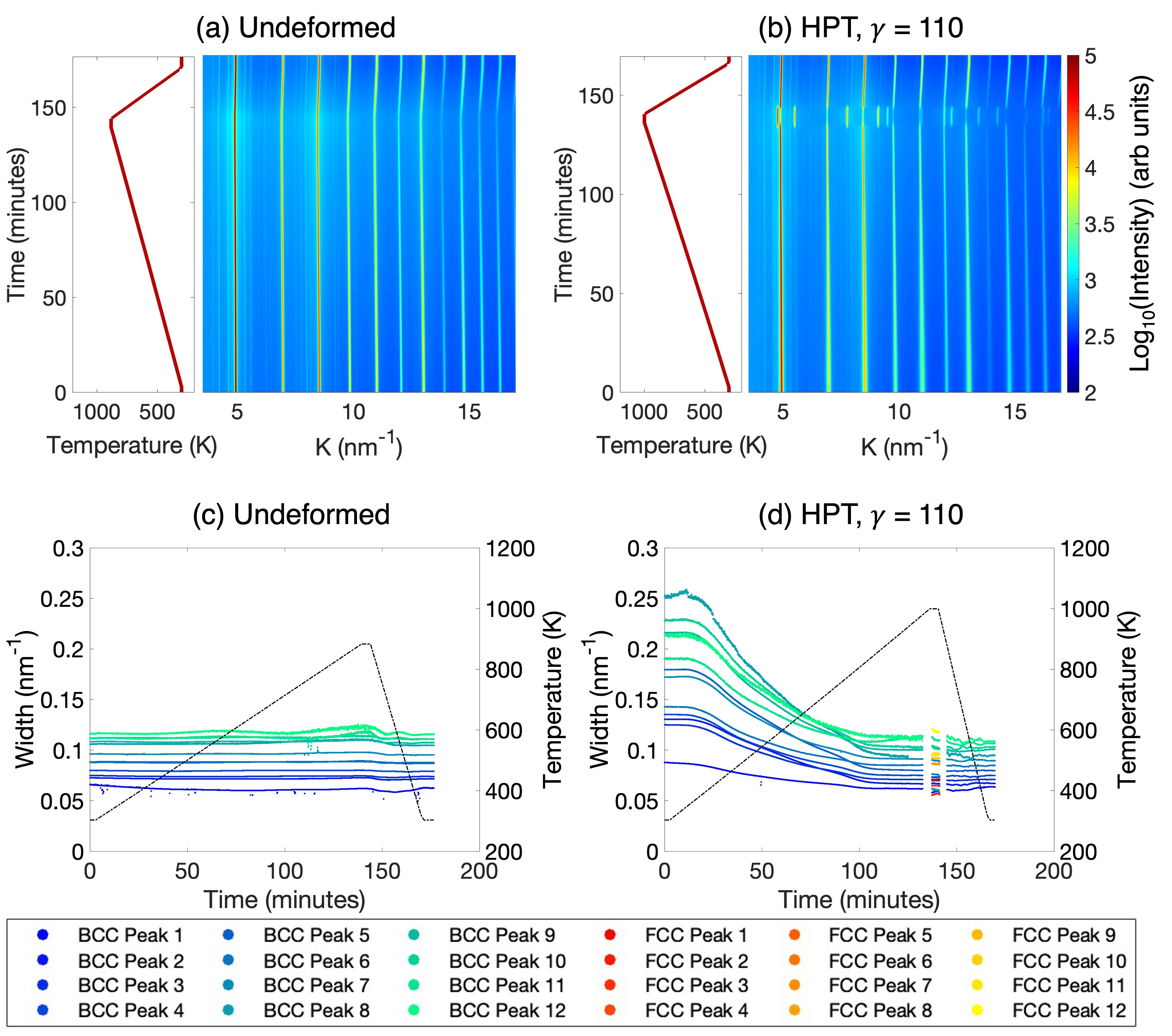}
			\caption{The temperature ramp and time vs. $K = 2\sin\theta/\lambda$ for the (a) undeformed and (b) HPT samples annealed to a lower target temperature compared to Figure \ref{fig:fastramp}. (c)-(d) Their respective peak widths (FWHM) as a function of time as well as the temperature profile (black dashed line). The ferrite/martensite phase was fitted to a body-centred cubic (BCC) structure, and the austenite was fitted to a face-centred cubic (FCC) structure. }
			\label{fig:slowramp}
		\end{figure}
		
		Another set of samples was annealed for this study but not discussed in the main text as the temperature ramp was not well-controlled (Figure \ref{fig:slowramp}). Specifically, for the annealing of the undeformed sample, the peak temperature only reached 883 K due to poor thermal contact. The peak widths of the undeformed sample did not change significantly during the whole annealing cycle and no phase transitions occurred. For the deformed sample, the peak annealing temperature was 1000 K, and it underwent a phase transformation starting at 977 K. The peak widths of the ferrite/martensite phase right before the phase transformation occurred are comparable to the peak widths of the undeformed sample. Upon cooling the austenite phase disappeared around 913 K. Due to the different heating rate of these samples compared to the samples presented in Figure \ref{fig:fastramp}, it is not surprising that the phase transition occurred at a different temperature. 
		
		The key difference compared to the first set of annealed samples (Figure \ref{fig:fastramp}) is that the peak widths following cooling remained small and there is no retained austenite phase (Figure \ref{fig:slowramp}). The continuity of the Debye-Scherrer rings (included in the supplementary files) after cooling is reduced compared to the initial state. This suggests that some grain growth has taken place to produce crystals that are dislocation- and strain-free, as expected. This gives further evidence that the retained austenite phase, and associated residual stress, is the cause of peak broadening post-annealing in the set of annealed samples discussed in Section \ref{sec:i12} and Figure \ref{fig:fastramp}.
		
	\section*{Supplementary Files}
		Four movies are included in the supplementary files which show the raw diffraction patterns from the in-situ annealing portion of the study. The samples are discussed alongside Figure \ref{fig:qtemp} (main text) and \ref{fig:fastramp} (Appendix E). The `LowTemp' files correspond to the samples discussed in Figure \ref{fig:slowramp} (Appendix E).

	\bibliographystyle{ieeetr}
	\bibliography{references}

\begin{thebibliography}{10}

\bibitem{Moslang2005}
A.~M{\"{o}}slang, E.~Diegele, M.~Klimiankou, R.~L{\"{a}}sser, R.~Lindau,
  E.~Lucon, E.~Materna-Morris, C.~Petersen, R.~Pippan, J.~W. Rensman, M.~Rieth,
  B.~{Van Der Schaaf}, H.~C. Schneider, and F.~Tavassoli, ``{Towards reduced
  activation structural materials data for fusion DEMO reactors},'' {\em
  Nuclear Fusion}, vol.~45, no.~7, p.~649, 2005.

\bibitem{Lindau2005}
R.~Lindau, A.~M{\"{o}}slang, M.~Rieth, M.~Klimiankou, E.~Materna-Morris,
  A.~Alamo, A.~A. Tavassoli, C.~Cayron, A.~M. Lancha, P.~Fernandez, N.~Baluc,
  R.~Sch{\"{a}}ublin, E.~Diegele, G.~Filacchioni, J.~W. Rensman, B.~V. Schaaf,
  E.~Lucon, and W.~Dietz, ``{Present development status of EUROFER and
  ODS-EUROFER for application in blanket concepts},'' {\em Fusion Engineering
  and Design}, vol.~75-79, no.~SUPPL., pp.~989--996, 2005.

\bibitem{Zinkle2012}
S.~J. Zinkle, ``{Radiation-induced effects on microstructure},'' in {\em
  Comprehensive Nuclear Materials}, vol.~1, pp.~65--98, Elsevier Ltd, 2012.

\bibitem{Ehrlich1999}
K.~Ehrlich, ``{The development of structural materials for fusion reactors},''
  {\em Philosophical Transactions of the Royal Society of London. Series A:
  Mathematical, Physical and Engineering Sciences}, vol.~357, no.~1752,
  pp.~595--623, 1999.

\bibitem{Klimenkov2011}
M.~Klimenkov, E.~Materna-Morris, and A.~M{\"{o}}slang, ``{Characterization of
  radiation induced defects in EUROFER 97 after neutron irradiation},'' {\em
  Journal of Nuclear Materials}, vol.~417, no.~1-3, pp.~124--126, 2011.

\bibitem{Enikeev2019}
N.~A. Enikeev, V.~K. Shamardin, and B.~Radiguet, ``{Radiation Tolerance of
  Ultrafine-Grained Materials Fabricated by Severe Plastic Deformation},'' {\em
  Materials Transactions}, vol.~60, no.~9, pp.~1723--1731, 2019.

\bibitem{Zhang2018}
X.~Zhang, K.~Hattar, Y.~Chen, L.~Shao, J.~Li, C.~Sun, K.~Yu, N.~Li, M.~L.
  Taheri, H.~Wang, J.~Wang, and M.~Nastasi, ``{Radiation damage in
  nanostructured materials},'' {\em Progress in Materials Science}, vol.~96,
  pp.~217--321, 2018.

\bibitem{Etienne2011}
A.~Etienne, B.~Radiguet, N.~J. Cunningham, G.~R. Odette, R.~Valiev, and
  P.~Pareige, ``{Comparison of radiation-induced segregation in
  ultrafine-grained and conventional 316 austenitic stainless steels},'' {\em
  Ultramicroscopy}, vol.~111, no.~6, pp.~659--663, 2011.

\bibitem{Radiguet2008}
B.~Radiguet, A.~Etienne, P.~Pareige, X.~Sauvage, and R.~Valiev, ``{Irradiation
  behavior of nanostructured 316 austenitic stainless steel},'' {\em Journal of
  Materials Science}, vol.~43, no.~23-24, pp.~7338--7343, 2008.

\bibitem{Mazilkin2020}
A.~Mazilkin, Y.~Ivanisenko, X.~Sauvage, A.~Etienne, B.~Radiguet, R.~Valiev,
  M.~Abramova, and N.~Enikeev, ``{Nanostructured Fe–Cr–W Steel Exhibits
  Enhanced Resistance to Self-Ion Irradiation},'' {\em Advanced Engineering
  Materials}, vol.~22, no.~10, p.~1901333, 2020.

\bibitem{Alsabbagh2014}
A.~Alsabbagh, A.~Sarkar, B.~Miller, J.~Burns, L.~Squires, D.~Porter, J.~I.
  Cole, and K.~Murty, ``{Microstructure and mechanical behavior of neutron
  irradiated ultrafine grained ferritic steel},'' {\em Materials Science and
  Engineering: A}, vol.~615, pp.~128--138, oct 2014.

\bibitem{Gigax2017}
J.~G. Gigax, H.~Kim, T.~Chen, F.~Garner, and L.~Shao, ``{Radiation instability
  of equal channel angular extruded T91 at ultra-high damage levels},'' {\em
  Acta Materialia}, vol.~132, pp.~395--404, 2017.

\bibitem{Song2014}
M.~Song, Y.~Wu, D.~Chen, X.~Wang, C.~Sun, K.~Yu, Y.~Chen, L.~Shao, Y.~Yang,
  K.~Hartwig, and X.~Zhang, ``{Response of equal channel angular extrusion
  processed ultrafine-grained T91 steel subjected to high temperature heavy ion
  irradiation},'' {\em Acta Materialia}, vol.~74, pp.~285--295, aug 2014.

\bibitem{Sun2015}
C.~Sun, S.~Zheng, C.~C. Wei, Y.~Wu, L.~Shao, Y.~Yang, K.~T. Hartwig, S.~A.
  Maloy, S.~J. Zinkle, T.~R. Allen, H.~Wang, and X.~Zhang, ``{Superior
  radiation-resistant nanoengineered austenitic 304L stainless steel for
  applications in extreme radiation environments},'' {\em Scientific Reports},
  vol.~5, p.~7801, jan 2015.

\bibitem{Aydogan2017}
E.~Aydogan, T.~Chen, J.~Gigax, D.~Chen, X.~Wang, P.~Dzhumaev, O.~Emelyanova,
  M.~Ganchenkova, B.~Kalin, M.~Leontiva-Smirnova, R.~Valiev, N.~Enikeev,
  M.~Abramova, Y.~Wu, W.~Lo, Y.~Yang, M.~Short, S.~Maloy, F.~Garner, and
  L.~Shao, ``{Effect of self-ion irradiation on the microstructural changes of
  alloy EK-181 in annealed and severely deformed conditions},'' {\em Journal of
  Nuclear Materials}, vol.~487, pp.~96--104, 2017.

\bibitem{Cao2018}
Y.~Cao, S.~Ni, X.~Liao, M.~Song, and Y.~Zhu, ``{Structural evolutions of
  metallic materials processed by severe plastic deformation},'' {\em Materials
  Science and Engineering: R: Reports}, vol.~133, pp.~1--59, 2018.

\bibitem{Edalati2022a}
P.~Edalati, A.~Mohammadi, M.~Ketabchi, and K.~Edalati, ``{Microstructure and
  microhardness of dual-phase high-entropy alloy by high-pressure torsion:
  Twins and stacking faults in FCC and dislocations in BCC},'' {\em Journal of
  Alloys and Compounds}, vol.~894, p.~162413, 2022.

\bibitem{Edalati2022}
K.~Edalati, A.~Bachmaier, V.~A. Beloshenko, Y.~Beygelzimer, V.~D. Blank, W.~J.
  Botta, K.~Bry{\l}a, J.~{\v{C}}{\'{i}}{\v{z}}ek, S.~Divinski, N.~A. Enikeev,
  Y.~Estrin, G.~Faraji, R.~B. Figueiredo, M.~Fuji, T.~Furuta, T.~Grosdidier,
  J.~Gubicza, A.~Hohenwarter, Z.~Horita, J.~Huot, Y.~Ikoma, M.~Jane{\v{c}}ek,
  M.~Kawasaki, P.~Kr{\'{a}}l, S.~Kuramoto, T.~G. Langdon, D.~R. Leiva, V.~I.
  Levitas, A.~Mazilkin, M.~Mito, H.~Miyamoto, T.~Nishizaki, R.~Pippan, V.~V.
  Popov, E.~N. Popova, G.~Purcek, O.~Renk, {\'{A}}.~R{\'{e}}v{\'{e}}sz,
  X.~Sauvage, V.~Sklenicka, W.~Skrotzki, B.~B. Straumal, S.~Suwas, L.~S. Toth,
  N.~Tsuji, R.~Z. Valiev, G.~Wilde, M.~J. Zehetbauer, and X.~Zhu,
  ``{Nanomaterials by severe plastic deformation: review of historical
  developments and recent advances},'' {\em Materials Research Letters},
  vol.~10, pp.~163--256, apr 2022.

\bibitem{Valiev2002}
R.~Z. Valiev, I.~V. Alexandrov, Y.~T. Zhu, and T.~C. Lowe, ``{Paradox of
  Strength and Ductility in Metals Processed Bysevere Plastic Deformation},''
  {\em Journal of Materials Research}, vol.~17, no.~1, pp.~5--8, 2002.

\bibitem{Zhilyaev2008}
A.~P. Zhilyaev and T.~G. Langdon, ``{Using high-pressure torsion for metal
  processing: Fundamentals and applications},'' {\em Progress in Materials
  Science}, vol.~53, no.~6, pp.~893--979, 2008.

\bibitem{BohlerBleche2003}
{B{\"{o}}hler Bleche}, ``{Certificate No. 201308 Datasheet},'' tech. rep.,
  V{\"{o}}lklingen, Germany, 2003.

\bibitem{Ziegler2010}
J.~F. Ziegler, M.~D. Ziegler, and J.~P. Biersack, ``{SRIM - The stopping and
  range of ions in matter (2010)},'' {\em Nuclear Instruments and Methods in
  Physics Research, Section B: Beam Interactions with Materials and Atoms},
  vol.~268, no.~11-12, pp.~1818--1823, 2010.

\bibitem{Derlet2020}
P.~M. Derlet and S.~L. Dudarev, ``{Microscopic structure of a heavily
  irradiated material},'' {\em Physical Review Materials}, vol.~4, no.~2,
  p.~023605, 2020.

\bibitem{Song2020}
K.~Song, S.~Das, A.~Reza, N.~W. Phillips, R.~Xu, H.~Yu, K.~Mizohata, D.~E.
  Armstrong, and F.~Hofmann, ``{Characterising Ion-Irradiated FeCr: Hardness,
  Thermal Diffusivity and Lattice Strain},'' {\em Acta Materialia}, vol.~201,
  pp.~535--546, 2020.

\bibitem{Song2023}
K.~Song, D.~Sheyfer, K.~Mizohata, M.~Zhang, W.~Liu, D.~G{\"{u}}rsoy, D.~Yang,
  I.~Tolkachev, H.~Yu, D.~E.~J. Armstrong, and F.~Hofmann, ``{Dose and
  compositional dependence of irradiation-induced property change in FeCr},''
  {\em arXiv preprint}, 2023.

\bibitem{Calcagnotto2010}
M.~Calcagnotto, D.~Ponge, E.~Demir, and D.~Raabe, ``{Orientation gradients and
  geometrically necessary dislocations in ultrafine grained dual-phase steels
  studied by 2D and 3D EBSD},'' {\em Materials Science and Engineering: A},
  vol.~527, no.~10-11, pp.~2738--2746, 2010.

\bibitem{Wright2015}
S.~I. Wright, D.~P. Field, and M.~M. Nowell, ``{Post processing effects on GND
  calculations from EBSD-based orientation measurements},'' {\em IOP Conference
  Series: Materials Science and Engineering}, vol.~89, no.~1, p.~012049, 2015.

\bibitem{Thompson2009}
S.~P. Thompson, J.~E. Parker, J.~Potter, T.~P. Hill, A.~Birt, T.~M. Cobb,
  F.~Yuan, and C.~C. Tang, ``{Beamline I11 at Diamond: A new instrument for
  high resolution powder diffraction},'' {\em Review of Scientific
  Instruments}, vol.~80, no.~7, 2009.

\bibitem{Tartoni2008}
N.~Tartoni, S.~P. Thompson, C.~C. Tang, B.~L. Willis, G.~E. Derbyshire, A.~G.
  Wright, S.~C. Jaye, J.~M. Homer, J.~D. Pizzey, and A.~M.~T. Bell,
  ``{High-performance X-ray detectors for the new powder diffraction beamline
  I11 at Diamond},'' {\em Journal of Synchrotron Radiation}, vol.~15, no.~1,
  pp.~43--49, 2008.

\bibitem{Filik2017}
J.~Filik, A.~W. Ashton, P.~C.~Y. Chang, P.~A. Chater, S.~J. Day,
  M.~Drakopoulos, M.~W. Gerring, M.~L. Hart, O.~V. Magdysyuk, S.~Michalik,
  A.~Smith, C.~C. Tang, N.~J. Terrill, M.~T. Wharmby, and H.~Wilhelm,
  ``{Processing two-dimensional X-ray diffraction and small-angle scattering
  data in DAWN 2},'' {\em J. Appl. Cryst}, vol.~50, pp.~959--966, 2017.

\bibitem{Paul2006}
A.~Pa{\'{u}}l, A.~Beirante, N.~Franco, E.~Alves, and J.~A. Odriozola, ``{Phase
  Transformation and Structural Studies of EUROFER RAFM Alloy},'' {\em
  Materials Science Forum}, vol.~514-516, pp.~500--504, may 2006.

\bibitem{Ribarik2019}
G.~Rib{\'{a}}rik, B.~J{\'{o}}ni, and T.~Ung{\'{a}}r, ``{Global optimum of
  microstructure parameters in the CMWP line-profile-analysis method by
  combining Marquardt-Levenberg and Monte-Carlo procedures},'' {\em Journal of
  Materials Science and Technology}, vol.~35, no.~7, pp.~1508--1514, 2019.

\bibitem{Ribarik2020}
G.~Rib{\'{a}}rik, B.~J{\'{o}}ni, and T.~Ung{\'{a}}r, ``{The Convolutional
  Multiple Whole Profile (CMWP) Fitting Method, a Global Optimization Procedure
  for Microstructure Determination},'' {\em Crystals}, vol.~10, no.~7, p.~623,
  2020.

\bibitem{Wilkens1970}
M.~Wilkens, ``{The determination of density and distribution of dislocations in
  deformed single crystals from broadened X-ray diffraction profiles},'' {\em
  Physica Status Solidi (a)}, vol.~2, no.~2, pp.~359--370, 1970.

\bibitem{Ungar1999a}
T.~Ung{\'{a}}r and G.~Tichy, ``{The effect of dislocation contrast on X-ray
  line profiles in untextured polycrystals},'' {\em Physica Status Solidi (A)
  Applied Research}, vol.~171, no.~2, pp.~425--434, 1999.

\bibitem{Reza2020b}
A.~Reza, C.~A. Dennett, M.~P. Short, J.~Waite, Y.~Zayachuk, C.~M. Magazzeni,
  S.~Hills, and F.~Hofmann, ``{Non-contact, non-destructive mapping of thermal
  diffusivity and surface acoustic wave speed using transient grating
  spectroscopy},'' {\em Review of Scientific Instruments}, vol.~91, no.~5,
  2020.

\bibitem{Kading1995}
O.~W. Kading, H.~Skurk, A.~A. Maznev, and E.~Matthias, ``{Transient thermal
  gratings at surfaces for thermal characterization of bulk materials and thin
  films},'' {\em Applied Physics A Materials Science and Processing}, vol.~61,
  no.~3, pp.~253--261, 1995.

\bibitem{Dennett2017}
C.~A. Dennett and M.~P. Short, ``{Time-resolved, dual heterodyne phase
  collection transient grating spectroscopy},'' {\em Applied Physics Letters},
  vol.~110, no.~21, p.~211106, 2017.

\bibitem{Ganeev2018}
A.~Ganeev, M.~Nikitina, V.~Sitdikov, R.~Islamgaliev, A.~Hoffman, and H.~Wen,
  ``{Effects of the Tempering and High-Pressure Torsion Temperatures on
  Microstructure of Ferritic/Martensitic Steel Grade 91},'' {\em Materials},
  vol.~11, no.~4, p.~627, 2018.

\bibitem{Arredondo2020}
R.~Arredondo, M.~Balden, A.~Mutzke, U.~von Toussaint, S.~Elgeti,
  T.~H{\"{o}}schen, K.~Schlueter, M.~Mayer, M.~Oberkofler, and W.~Jacob,
  ``{Impact of surface enrichment and morphology on sputtering of EUROFER by
  deuterium},'' {\em Nuclear Materials and Energy}, vol.~23, p.~100749, 2020.

\bibitem{Hafok2008}
M.~Hafok and R.~Pippan, ``{High-pressure torsion applied to nickel single
  crystals},'' {\em Philosophical Magazine}, vol.~88, no.~12, pp.~1857--1877,
  2008.

\bibitem{Pippan2010}
R.~Pippan, S.~Scheriau, A.~Taylor, M.~Hafok, A.~Hohenwarter, and A.~Bachmaier,
  ``{Saturation of Fragmentation During Severe Plastic Deformation},'' {\em
  Annual Review of Materials Research}, vol.~40, no.~1, pp.~319--343, 2010.

\bibitem{Rathmayr2013}
G.~B. Rathmayr, A.~Hohenwarter, and R.~Pippan, ``{Influence of grain shape and
  orientation on the mechanical properties of high pressure torsion deformed
  nickel},'' {\em Materials Science and Engineering: A}, vol.~560,
  pp.~224--231, 2013.

\bibitem{Naghdy2017}
S.~Naghdy, H.~Pirgazi, P.~Verleysen, R.~Petrov, and L.~Kestens,
  ``{Morphological and crystallographic anisotropy of severely deformed
  commercially pure aluminium by three-dimensional electron backscatter
  diffraction},'' {\em Journal of Applied Crystallography}, vol.~50, no.~5,
  pp.~1512--1523, 2017.

\bibitem{Raabe2014}
D.~Raabe, ``{Recovery and Recrystallization: Phenomena, Physics, Models,
  Simulation},'' in {\em Physical Metallurgy}, pp.~2291--2397, Elsevier, 2014.

\bibitem{Yao2008}
Z.~Yao, M.~Hern{\'{a}}ndez-Mayoral, M.~L. Jenkins, and M.~A. Kirk, ``{Heavy-ion
  irradiations of Fe and Fe–Cr model alloys Part 1: Damage evolution in
  thin-foils at lower doses},'' {\em Philosophical Magazine}, vol.~88, no.~21,
  pp.~2851--2880, 2008.

\bibitem{Baczynski1996}
J.~Baczynski and J.~J. Jonas, ``{Texture development during the torsion testing
  of $\alpha$-iron and two IF steels},'' {\em Acta Materialia}, vol.~44,
  no.~11, pp.~4273--4288, 1996.

\bibitem{Ungar1996}
T.~Ung{\'{a}}r and A.~Borb{\'{e}}ly, ``{The effect of dislocation contrast on
  x-ray line broadening: A new approach to line profile analysis},'' {\em
  Applied Physics Letters}, vol.~69, no.~21, pp.~3173--3175, 1996.

\bibitem{Borbely2022}
A.~Borb{\'{e}}ly, ``{The modified Williamson-Hall plot and dislocation density
  evaluation from diffraction peaks},'' {\em Scripta Materialia}, vol.~217,
  p.~114768, 2022.

\bibitem{Tyralis2013}
H.~Tyralis, D.~Koutsoyiannis, and S.~Kozanis, ``{An algorithm to construct
  Monte Carlo confidence intervals for an arbitrary function of probability
  distribution parameters},'' {\em Computational Statistics}, vol.~28, no.~4,
  pp.~1501--1527, 2013.

\bibitem{Ungar2005}
T.~Ung{\'{a}}r, G.~Tichy, J.~Gubicza, and R.~J. Hellmig, ``{Correlation between
  subgrains and coherently scattering domains},'' {\em Powder Diffraction},
  vol.~20, no.~4, pp.~366--375, 2005.

\bibitem{Li2018}
X.~Li, X.~Li, S.~Sch{\"{o}}necker, R.~Li, J.~Zhao, and L.~Vitos,
  ``{Understanding the mechanical properties of reduced activation steels},''
  {\em Materials and Design}, vol.~146, pp.~260--272, 2018.

\bibitem{Danon2002}
A.~Dan{\'{o}}n and A.~Alamo, ``{Behavior of Eurofer97 reduced activation
  martensitic steel upon heating and continuous cooling},'' {\em Journal of
  Nuclear Materials}, vol.~307-311, no.~1 SUPPL., pp.~479--483, 2002.

\bibitem{Kumar2021}
D.~Kumar, J.~Hargreaves, A.~Bharj, A.~Scorror, L.~M. Harding,
  H.~Dominguez-Andrade, R.~Holmes, R.~Burrows, H.~Dawson, A.~D. Warren, P.~E.
  Flewitt, and T.~L. Martin, ``{The effects of fusion reactor thermal
  transients on the microstructure of Eurofer-97 steel},'' {\em Journal of
  Nuclear Materials}, vol.~554, p.~153084, 2021.

\bibitem{Hidalgo2017}
J.~Hidalgo, K.~Findley, and M.~Santofimia, ``{Thermal and mechanical stability
  of retained austenite surrounded by martensite with different degrees of
  tempering},'' {\em Materials Science and Engineering: A}, vol.~690,
  pp.~337--347, apr 2017.

\bibitem{Tirumalasetty2012}
G.~K. Tirumalasetty, M.~A. {Van Huis}, C.~Kwakernaak, J.~Sietsma, W.~G. Sloof,
  and H.~W. Zandbergen, ``{Deformation-induced austenite grain rotation and
  transformation in TRIP-assisted steel},'' {\em Acta Materialia}, vol.~60,
  no.~3, pp.~1311--1321, 2012.

\bibitem{Nakada2016}
N.~Nakada, Y.~Ishibashi, T.~Tsuchiyama, and S.~Takaki, ``{Self-stabilization of
  untransformed austenite by hydrostatic pressure via martensitic
  transformation},'' {\em Acta Materialia}, vol.~110, pp.~95--102, 2016.

\bibitem{Villa2012}
M.~Villa, F.~B. Grumsen, K.~Pantleon, and M.~A. Somers, ``{Martensitic
  transformation and stress partitioning in a high-carbon steel},'' {\em
  Scripta Materialia}, vol.~67, no.~6, pp.~621--624, 2012.

\bibitem{Mergia2008}
K.~Mergia and N.~Boukos, ``{Structural, thermal, electrical and magnetic
  properties of Eurofer 97 steel},'' {\em Journal of Nuclear Materials},
  vol.~373, no.~1-3, pp.~1--8, 2008.

\bibitem{Graham2021}
M.~W. Graham, J.~C. King, T.~R. Pavlov, C.~A. Adkins, S.~C. Middlemas, and
  D.~P. Guillen, ``{Impact of neutron irradiation on the thermophysical
  properties of additively manufactured stainless steel and inconel},'' {\em
  Journal of Nuclear Materials}, vol.~549, p.~152861, 2021.

\bibitem{Wylie2022}
A.~P.~C. Wylie, K.~B. Woller, S.~A.~A. {Al Dajani}, B.~R. Dacus, E.~J.
  Pickering, M.~Preuss, and M.~P. Short, ``{Thermal diffusivity in
  ion-irradiated single-crystal iron, chromium, vanadium, and tungsten measured
  using transient grating spectroscopy},'' {\em Journal of Applied Physics},
  vol.~132, no.~4, p.~045102, 2022.

\bibitem{Badry2020}
F.~Badry and K.~Ahmed, ``{A new model for the effective thermal conductivity of
  polycrystalline solids},'' {\em AIP Advances}, vol.~10, no.~10, p.~105021,
  2020.

\bibitem{Dong2014}
H.~Dong, B.~Wen, and R.~Melnik, ``{Relative importance of grain boundaries and
  size effects in thermal conductivity of nanocrystalline materials},'' {\em
  Scientific Reports}, vol.~4, no.~1, p.~7037, 2014.

\bibitem{Hofmann2019}
F.~Hofmann, M.~P. Short, and C.~A. Dennett, ``{Transient grating spectroscopy:
  An ultrarapid, nondestructive materials evaluation technique},'' {\em MRS
  Bulletin}, vol.~44, no.~5, pp.~392--402, 2019.

\bibitem{Hofmann2015}
F.~Hofmann, D.~Nguyen-Manh, M.~Gilbert, C.~Beck, J.~Eliason, A.~Maznev, W.~Liu,
  D.~Armstrong, K.~Nelson, and S.~Dudarev, ``{Lattice swelling and modulus
  change in a helium-implanted tungsten alloy: X-ray micro-diffraction, surface
  acoustic wave measurements, and multiscale modelling},'' {\em Acta
  Materialia}, vol.~89, pp.~352--363, 2015.

\bibitem{Duncan2016}
R.~A. Duncan, F.~Hofmann, A.~Vega-Flick, J.~K. Eliason, A.~A. Maznev, A.~G.
  Every, and K.~A. Nelson, ``{Increase in elastic anisotropy of single crystal
  tungsten upon He-ion implantation measured with laser-generated surface
  acoustic waves},'' {\em Applied Physics Letters}, vol.~109, no.~15,
  p.~151906, 2016.

\bibitem{Dennett2018}
C.~Dennett, K.~So, A.~Kushima, D.~Buller, K.~Hattar, and M.~Short, ``{Detecting
  self-ion irradiation-induced void swelling in pure copper using transient
  grating spectroscopy},'' {\em Acta Materialia}, vol.~145, pp.~496--503, 2018.

\bibitem{Dennett2019}
C.~A. Dennett, D.~L. Buller, K.~Hattar, and M.~P. Short, ``{Real-time
  thermomechanical property monitoring during ion beam irradiation using in
  situ transient grating spectroscopy},'' {\em Nuclear Instruments and Methods
  in Physics Research Section B: Beam Interactions with Materials and Atoms},
  vol.~440, pp.~126--138, 2019.

\bibitem{Strangward-Pryce2023}
G.~Strangward-Pryce, K.~Song, K.~Mizohata, and F.~Hofmann, ``{The Effect of
  High-Pressure Torsion on Irradiation Hardening of Eurofer-97},'' {\em Nuclear
  Materials and Energy}, p.~101468, 2023.

\bibitem{Zhilyaev2003}
A.~P. Zhilyaev, J.~Gubicza, G.~Nurislamova, {\'{A}}.~R{\'{e}}v{\'{e}}sz,
  S.~Suri{\~{n}}ach, M.~D. Bar{\'{o}}, and T.~Ung{\'{a}}r, ``{Microstructural
  characterization of ultrafine-grained nickel},'' {\em physica status solidi
  (a)}, vol.~198, no.~2, pp.~263--271, 2003.

\bibitem{Gubicza2004}
J.~Gubicza, N.~Chinh, Z.~Horita, and T.~Langdon, ``{Effect of Mg addition on
  microstructure and mechanical properties of aluminum},'' {\em Materials
  Science and Engineering: A}, vol.~387-389, pp.~55--59, 2004.

\bibitem{Gallet2023}
J.~Gallet, M.~Perez, R.~Guillou, C.~Ernould, C.~{Le Bourlot}, C.~Langlois,
  B.~Beausir, E.~Bouzy, T.~Chaise, and S.~Cazottes, ``{Experimental measurement
  of dislocation density in metallic materials: A quantitative comparison
  between measurements techniques (XRD, R-ECCI, HR-EBSD, TEM)},'' {\em
  Materials Characterization}, vol.~199, p.~112842, 2023.

\bibitem{Zhang2020}
Z.~Zhang, {\'{E}}.~{\'{O}}dor, D.~Farkas, B.~J{\'{o}}ni, G.~Rib{\'{a}}rik,
  G.~Tichy, S.~H. Nandam, J.~Ivanisenko, M.~Preuss, and T.~Ung{\'{a}}r,
  ``{Dislocations in Grain Boundary Regions: The Origin of Heterogeneous
  Microstrains in Nanocrystalline Materials},'' {\em Metallurgical and
  Materials Transactions A: Physical Metallurgy and Materials Science},
  vol.~51, no.~1, pp.~513--530, 2020.

\bibitem{Valiev2000}
R.~Z. Valiev, R.~K. Islamgaliev, and I.~V. Alexandrov, ``{Bulk nanostructured
  materials from severe plastic deformation},'' {\em Progress in Materials
  Science}, vol.~45, no.~2, pp.~103--189, 2000.

\bibitem{Valiev1993}
R.~Z. Valiev, A.~V. Korznikov, and R.~R. Mulyukov, ``{Structure and properties
  of ultrafine-grained materials produced by severe plastic deformation},''
  {\em Materials Science and Engineering: A}, vol.~168, no.~2, pp.~141--148,
  1993.

\bibitem{Jiang2013}
J.~Jiang, T.~B. Britton, and A.~J. Wilkinson, ``{Measurement of geometrically
  necessary dislocation density with high resolution electron backscatter
  diffraction: Effects of detector binning and step size},'' {\em
  Ultramicroscopy}, vol.~125, pp.~1--9, 2013.

\bibitem{Vitek2011}
V.~Vitek, ``{Atomic level computer modelling of crystal defects with emphasis
  on dislocations: Past, present and future},'' {\em Progress in Materials
  Science}, vol.~56, no.~6, pp.~577--585, 2011.

\bibitem{Duesbery1998}
M.~Duesbery and W.~Xu, ``{The motion of edge dislocations in body-centered
  cubic metals},'' {\em Scripta Materialia}, vol.~39, no.~3, pp.~283--287,
  1998.

\bibitem{Wei2006}
Q.~Wei, H.~Zhang, B.~Schuster, K.~Ramesh, R.~Valiev, L.~Kesckes, R.~Ddowding,
  L.~Magness, and K.~Cho, ``{Microstructure and mechanical properties of
  super-strong nanocrystalline tungsten processed by high-pressure torsion},''
  {\em Acta Materialia}, vol.~54, no.~15, pp.~4079--4089, 2006.

\bibitem{Cheng2013}
G.~Cheng, W.~Xu, W.~Jian, H.~Yuan, M.~Tsai, Y.~Zhu, Y.~Zhang, and P.~Millett,
  ``{Dislocations with edge components in nanocrystalline bcc Mo},'' {\em
  Journal of Materials Research}, vol.~28, no.~13, pp.~1820--1826, 2013.

\bibitem{Cheng2013a}
G.~Cheng, W.~Jian, W.~Xu, H.~Yuan, P.~Millett, and Y.~Zhu, ``{Grain Size Effect
  on Deformation Mechanisms of Nanocrystalline bcc Metals},'' {\em Materials
  Research Letters}, vol.~1, no.~1, pp.~26--31, 2013.

\bibitem{Wei2011}
Q.~Wei, Z.~Pan, X.~Wu, B.~Schuster, L.~Kecskes, and R.~Valiev,
  ``{Microstructure and mechanical properties at different length scales and
  strain rates of nanocrystalline tantalum produced by high-pressure
  torsion},'' {\em Acta Materialia}, vol.~59, no.~6, pp.~2423--2436, 2011.

\bibitem{Cui2016}
Y.~Cui, G.~Po, and N.~Ghoniem, ``{Temperature insensitivity of the flow stress
  in body-centered cubic micropillar crystals},'' {\em Acta Materialia},
  vol.~108, pp.~128--137, 2016.

\bibitem{Hull2011}
D.~Hull and D.~Bacon, ``{Jogs and the Intersection of Dislocations},'' in {\em
  Introduction to Dislocations}, pp.~137--155, Elsevier, 2011.

\bibitem{Huang2011}
L.~Huang, Q.-J. Li, Z.-W. Shan, J.~Li, J.~Sun, and E.~Ma, ``{A new regime for
  mechanical annealing and strong sample-size strengthening in body centred
  cubic molybdenum},'' {\em Nature Communications}, vol.~2, no.~1, p.~547,
  2011.

\bibitem{Humphreys1996}
F.~J. Humphreys and H.~M. Chan, ``{Discontinuous and continuous annealing
  phenomena in aluminium–nickel alloy},'' {\em Materials Science and
  Technology}, vol.~12, no.~2, pp.~143--148, 1996.

\bibitem{Hao2014}
T.~Hao, Z.~Q. Fan, S.~X. Zhao, G.~N. Luo, C.~S. Liu, and Q.~F. Fang,
  ``{Strengthening mechanism and thermal stability of severely deformed
  ferritic/martensitic steel},'' {\em Materials Science and Engineering: A},
  vol.~596, pp.~244--249, 2014.

\bibitem{El-Tahawy2017}
M.~El-Tahawy, Y.~Huang, H.~Choi, H.~Choe, J.~L. L{\'{a}}b{\'{a}}r, T.~G.
  Langdon, and J.~Gubicza, ``{High temperature thermal stability of
  nanocrystalline 316L stainless steel processed by high-pressure torsion},''
  {\em Materials Science and Engineering: A}, vol.~682, pp.~323--331, 2017.

\bibitem{JavadzadehKalahroudi2019}
F.~{Javadzadeh Kalahroudi}, H.~Koohdar, H.~R. Jafarian, Y.~Haung, T.~G.
  Langdon, and M.~Nili-Ahmadabadi, ``{On the microstructure and mechanical
  properties of an Fe-10Ni-7Mn martensitic steel processed by high-pressure
  torsion},'' {\em Materials Science and Engineering: A}, vol.~749, pp.~27--34,
  2019.

\bibitem{Reza2020}
A.~Reza, H.~Yu, K.~Mizohata, and F.~Hofmann, ``{Thermal diffusivity degradation
  and point defect density in self-ion implanted tungsten},'' {\em Acta
  Materialia}, vol.~193, pp.~270--279, 2020.

\bibitem{Rieth2003}
M.~Rieth, M.~Schirra, A.~Falkenstein, P.~Graf, S.~Heger, H.~Kempe, R.~Lindau,
  and H.~Zimmermann, ``{EUROFER 97. Tensile, charpy, creep and structural
  tests},'' tech. rep., Forschungszentrum Karlsruhe GmbH Technik und Umwelt,
  Karlsruhe, 2003.

\bibitem{Ungar2021}
T.~Ung{\'{a}}r, G.~Rib{\'{a}}rik, M.~Topping, R.~M.~A. Jones, X.~{Dan Xu},
  R.~Hulse, A.~Harte, G.~Tichy, C.~P. Race, P.~Frankel, and M.~Preuss,
  ``{Characterizing dislocation loops in irradiated polycrystalline Zr alloys
  by X-ray line profile analysis of powder diffraction patterns with
  satellites},'' {\em Journal of Applied Crystallography}, vol.~54, no.~3,
  pp.~803--821, 2021.

\bibitem{Balogh2016}
L.~Balogh, F.~Long, and M.~R. Daymond, ``{Contrast factors of
  irradiation-induced dislocation loops in hexagonal materials},'' {\em Journal
  of Applied Crystallography}, vol.~49, no.~6, pp.~2184--2200, 2016.

\end{thebibliography}

\end{document}